\documentclass[journal]{IEEEtran}

\usepackage{xcolor,soul,framed} 

\colorlet{shadecolor}{yellow}
\usepackage[pdftex]{graphicx}
\graphicspath{{../pdf/}{../jpeg/}}
\DeclareGraphicsExtensions{.pdf,.jpeg,.png}

\usepackage[cmex10]{amsmath}
\usepackage{array}
\usepackage{mdwmath}
\usepackage{mdwtab}
\usepackage{booktabs}
\usepackage{tabularx}
\usepackage{ragged2e}
\usepackage{eqparbox}
\usepackage{url}
\usepackage{algorithm}
\usepackage{algorithmicx}
\usepackage{algcompatible}
\usepackage{mathrsfs}
\usepackage{amssymb} 
\usepackage{mathtools}
\usepackage{mathrsfs} 
\usepackage{dutchcal}
\usepackage{booktabs}
\usepackage{url}
\usepackage{amsmath}
\usepackage{multirow}
\usepackage{breakurl}
\usepackage{enumerate}
\usepackage[hidelinks]{hyperref}
\usepackage{array}
\hypersetup{
	pdfauthor={Farid Zaredar, Morteza Amini},     
	pdftitle={A Collusion-Resistance Privacy-Preserving Aggregation Smart Metering Protocol for Operational Utility}, 
	pdfsubject={Smart Metering Privacy},   
	pdfkeywords={Smart Grid Privacy, Smart Meter Privacy,  Privacy-Preserving Collusion-Resistance Smart Metering Protocol, Partially Additive Homomorphic Encryption, Data Perturbation Technique}, 
	pdfcreator={LaTeX with hyperref}, 
	pdfproducer={pdflatex},     
	bookmarks=true,            
	bookmarksopen=true,        
	bookmarksnumbered=true,    
	colorlinks=false,          
}

\begin{document}
	\bstctlcite{IEEEexample:BSTcontrol}
	\title{A Collusion-Resistance Privacy-Preserving Smart Metering Protocol for Operational Utility}
	\author{Farid~Zaredar, and  
		Morteza~Amini
		
		\thanks{F. Zaredar is with the Data and Network Security Laboratory (DNSL), Department of Computer Engineering, Sharif University of Technology, Tehran, Iran (e-mail: farid.zaredar78@sharif.edu).}
	
		\thanks{M. Amini is with Department of Computer Engineering, Sharif University of Technology, Tehran, Iran (e-mail: amini@sharif.edu).}

}

	\markboth{}{Roberg \MakeLowercase{\textit{et al.}}: High-Efficiency Diode and Transistor Rectifiers}

	\maketitle

	\begin{abstract}
		Modern smart grids rely on advanced metering infrastructure (AMI) to collect
		fine-grained consumption readings for operational services such as grid
		monitoring, load forecasting, and demand--supply balancing. However, these
		high-frequency readings can reveal sensitive information about consumers'
		daily activities. To address this privacy concern, we propose a
		collusion-resistant privacy-preserving aggregation protocol for smart
		metering operational services. The protocol distributes noise-cancellation
		responsibility among a configurable group of \(K\) designated smart meters.
		Each non-designated meter perturbs its reading using \(K\) independent noise
		components, while the corresponding cancellation values ensure that the
		noise is removed only from the final aggregate.
		The protocol combines Paillier homomorphic encryption with a
		KEM--KDF--AEAD construction. Paillier encryption enables the aggregator to
		compute an encrypted aggregate without decrypting individual contributions,
		while authenticated encryption protects the exchanged noise components
		between smart meters. Under the considered collusion and meter-exposure model, the exact reading of a trusted and unexposed meter remains protected as long as at least one designated and one non-designated meter remain unexposed.
		We evaluate the protocol in terms of computational, memory, communication,
		and privacy overheads. Privacy is evaluated using normalized conditional
		entropy (NCE) and normalized root-mean-square error (NRMSE). The results show
		that increasing the noise scale increases NCE and therefore the uncertainty
		about individual readings, while the NRMSE evaluation quantifies the gradual
		loss of privacy as additional opposite-role meters are exposed. Overall, the
		proposed protocol provides exact aggregate consumption values required for
		operational services while protecting individual fine-grained readings
		against the considered adversarial coalition.
	\end{abstract}
	
	\begin{IEEEkeywords}
		Smart Grid, Smart Meter, Privacy,  Homomorphic Encryption, Data Perturbation
	\end{IEEEkeywords}

	%
	\IEEEpeerreviewmaketitle


	\section{Introduction}
	\IEEEPARstart{T}{he} digital evolution of energy infrastructure has introduced numerous advantages, including real-time monitoring and grid management, demand-supply balancing, self-healing, efficient energy generation and transmission, improved load forecasting, and rapid outage detection. Smart grids provide greater efficiency, strong reliability and sustainability, and better flexibility in electricity management compared to the traditional electrical grids, \cite{giaconi2018privacy, kumar2019smart}. Due to the inability of traditional power grids to meet electricity demand in the 21st century \cite{giaconi2018privacy}, many countries have adopted smart grids as their modernized power infrastructure. According to the U.S. Energy Information Administration (EIA), global energy consumption is projected to rise by 48\% between 2012 to 2040 \cite{EIA2016}. Additionally, the seamless integration of smart grids with renewable energy resources reduces cost and energy waste \cite{mazhar2023analysis}. This integration also contributes to lowering carbon emissions \cite{kua2023privacy}, thereby mitigating air pollution. In Mexico and Central America, considerable efforts have been made to increase electricity generation from renewable energy resources, such as solar and wind energy \cite{moreno2021comprehensive}. More specifically, Mexico and Central America aim to produce 50\% of their electricity from renewable energy resources by 2030 \cite{Ferreira2016}. Similar developments are evident in other regions: in Europe, smart-grid modernization and smart-metering infrastructure facilitate the integration of renewable energy resources, while in Japan, smart meters have become part of the electricity-system infrastructure and are complemented by IoT-enabled distributed energy resources that support further renewable-energy integration \cite{depaola2023smartgrids,meti2024energywhitepaper}.

	 Advanced metering infrastructure (AMI) enables a bidirectional flow of energy and information between smart meters and the metering data management system (MDMS) provided by the energy supplier \cite{ben2022privacy}. The AMI is superior to the previous technology, automated meter reading (AMR), which only supports unidirectional communication \cite{ansari2022state}. The AMI consists of three essential components: (1) smart meters, (2) communication networks, and (3) metering data management systems \cite{ansari2022state}. Smart meters report fine-grained consumption values at high frequency (i.e., every 15 minutes) to the utility provider. Communication technologies such as power-line communication (PLC) and radio access networks (RANs) facilitate the communication between intelligent meters and the MDMS. The MDMS collects, processes, and analyzes these fine-grained readings for various purposes such as billing, operational utilities and, different types of value-added services. 
	
\begin{center}
		\begin{figure}[htp]
			\includegraphics[width=3.5in]{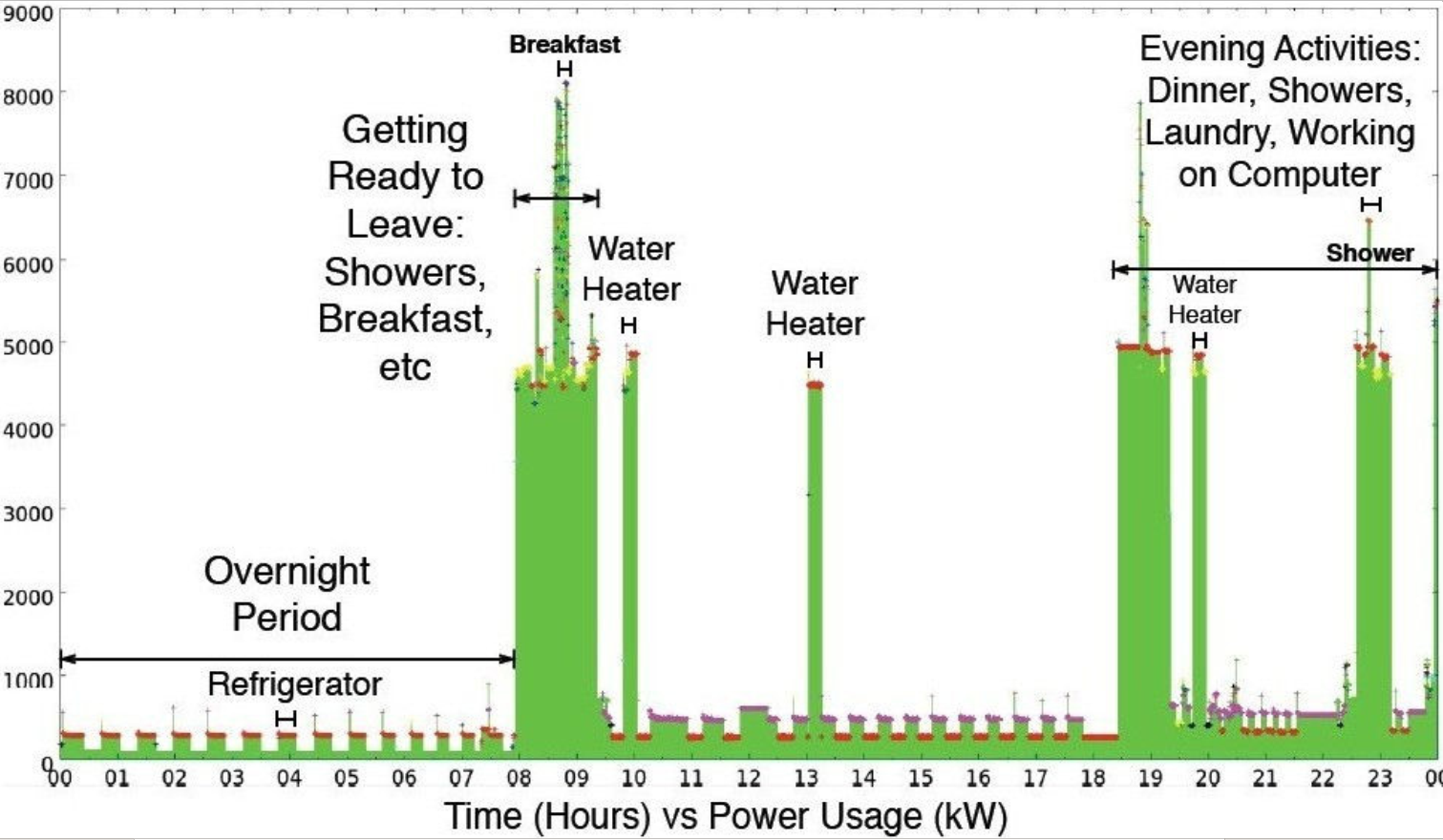}
			\caption{An example of a full-day power consumption trace \cite{souri2014smart}}
			\label{op-power-trace}
		\end{figure}
	\end{center}

	Although the collection of fine-grained consumption values is useful for grid management and monitoring, it also raises significant privacy concerns. For instance, with such data, the utility provider can infer a customer's daily life patterns (e.g., presence or absence, watching TV, playing video games, sleeping patterns, and other daily activities),  political orientations, types of home appliances  \cite{athanasiadis2021real, de2018home, halim2025nons} and even religious beliefs. Indeed, conducting an in-depth analysis of consumption values provides profound insight relevant to individuals' daily habits \cite{finster2015privacy}. Consequently, this can lead to serious privacy violations. Fig. \ref{op-power-trace} presents a sample power trace derived from a smart meter's daily consumption data. It highlights overnight period, breakfast time, office hours, and evening activities (e.g., taking shower, doing laundry, and working on a computer). Furthermore, more advanced data mining techniques can reveal more detailed information about the user \cite{souri2014smart}. 
	
	To address these privacy concerns, various privacy-preserving protocols have been introduced. Previous protocols have adopted different privacy-enhancing techniques, including data masking schemes, zero-knowledge proof, the use of trusted third parties, anonymization, secure multi-party computation, and bi-homomorphic encryptions. While these schemes offer security and privacy, they also introduce challenges related to implementation and deployment complexity, high computational and communication costs, and scalability issues.  
	
		In this paper, we propose a collusion-resistant privacy-preserving
		aggregation protocol for smart-metering operational services. The protocol
		enables smart meters to report their consumption at predefined intervals
		(e.g., every 15 minutes), while the utility provider obtains only the exact
		aggregate required for grid management, monitoring, and load forecasting.
		To protect individual readings, each non-designated smart meter perturbs its
		reading using \(K\) independent noise components, while \(K\) designated
		smart meters provide the corresponding cancellation values. The designated
		set changes across reporting intervals, thereby distributing the
		noise-cancellation responsibility among participating meters.
		
		The protocol combines Paillier homomorphic encryption with a hybrid construction based on a key encapsulation mechanism (KEM), key derivation function (KDF), and authenticated encryption with associated data (AEAD). Paillier encryption protects the submitted meter contributions and enables the aggregator to compute the encrypted aggregate without decrypting them, while the KEM–KDF–AEAD construction establishes pairwise shared secrets, derives fresh interval-specific keys, and provides end-to-end authenticated protection for the individual noise components exchanged between the corresponding smart meters. The protocol is designed to protect the exact individual readings of trusted and unexposed meters even when the aggregator and utility provider collude and combine their protocol views with the internal states of a limited number of exposed meters.
		
		Our main contributions can be summarized as follows:
		\begin{enumerate}
			\item We propose a privacy-preserving aggregation protocol that provides
			the utility provider with the exact aggregate consumption value required
			for operational services without revealing the exact individual reading
			of a trusted and unexposed smart meter.
			
			\item We introduce a configurable group of \(K\) designated smart meters
			that distributes the noise-cancellation responsibility. Each
			non-designated meter generates \(K\) independent noise components whose
			total variance remains independent of \(K\), while the corresponding
			cancellation values preserve the exact aggregate.
			
			\item We provide resistance against collusion between the aggregator and
			the utility provider in the presence of limited meter exposure. Provided that at least one designated and one non-designated meter remain unexposed, the adversarial coalition cannot exactly recover the reading of a trusted and unexposed meter.
			
			\item We employ a hybrid cryptographic construction in which each meter
			submits only one Paillier-encrypted consumption contribution per
			reporting attempt, while lightweight KEM--KDF--AEAD protection is used
			for the exchanged noise components. We evaluate the resulting
			computational, memory, communication, and privacy overheads.
		\end{enumerate}
		
		The remainder of this paper is organized as follows: Section II reviews
		related work. Section III presents the system model, threat model,
		assumptions, and design goals. Section IV describes the proposed protocol.
		Section V evaluates its performance and privacy and presents the security
		analysis. Finally, Section VI concludes the paper.

	
	\section{Related Work}
	
	In the literature, various privacy-preserving techniques have been proposed to support smart-grid operational services such as grid monitoring, load forecasting, and demand--supply balancing. These services require access to high-frequency consumption information, while direct disclosure of fine-grained readings may reveal customers' daily activities and consumption patterns. Existing schemes address this conflict through encrypted aggregation, secret sharing, secure multi-party computation, data perturbation, anonymization, and distributed masking. For the purpose of this work, we classify the most relevant approaches into four categories: (1) aggregation via third parties, (2) aggregation without trusted third parties, (3) anonymous reporting via third parties, and (4) reporting via anonymous overlay networks.
	
	\subsection{Aggregation via Third Parties}
	
	In this class, intermediary nodes such as gateways, aggregators, substations, or fog nodes collect and process consumption values before forwarding aggregated information to the utility provider. Although such architectures can reduce the processing burden on the control center, privacy may depend on the trust assumptions associated with the intermediary.
	
	Bohli et al. \cite{bohli2010privacy} present two privacy-preserving approaches: one relies on a fully trusted third party that collects individual meter readings and reports only their aggregate, while the other perturbs consumption readings to reduce the utility provider's certainty about individual values. Garcia et al. \cite{garcia2011privacy} combine additive homomorphic encryption with additive secret sharing so that only an aggregate is revealed; however, their privacy guarantees weaken when too few meters participate or multiple devices are compromised. Lu et al. \cite{lu2012eppa} employ Paillier encryption with a super-increasing sequence for encrypted multidimensional aggregation, whereas Molina-Markham et al. \cite{molina2010private} rely on a trusted aggregator to conceal meter identities from an honest-but-curious utility provider.
	
	Other approaches extend intermediary-assisted aggregation using secure computation, perturbation, and multidimensional processing. Mustafa et al. \cite{mustafa2019secure} employ secure multi-party computation in which meter measurements are divided into secret shares and aggregated by data communication company servers. Wang et al. \cite{wang2022preen} combine differential privacy with ElGamal-based re-encryption, while Wang et al. \cite{wang2023privacy} use additive ElGamal encryption and binary encoding for multidimensional aggregation. Li et al. \cite{li2023fine} combine Paillier encryption, bilinear pairing, and binary encoding to support different aggregation classes. Zhang et al. \cite{zhang2020privacy} instead employ a dedicated operational channel in which a trusted substation aggregates high-frequency readings before reporting them to the network operator.
	
	More recent schemes consider stronger adversarial models and additional operational requirements. Liu et al. \cite{liu2023certificateless} propose a certificateless multidimensional aggregation scheme based on Paillier encryption and key negotiation in a fog-computing architecture. Their scheme is designed to resist collusion between the control center and fog nodes and also supports fault tolerance and dynamic membership. Li et al. \cite{li2026certificateless} further combine the Chinese Remainder Theorem (CRT), updated blinding factors, and Paillier homomorphic encryption to provide certificateless multidimensional aggregation with fault tolerance and resistance to collusion attacks.
	
	Naeem et al. \cite{naeem2026dalbs} propose DALBS, which combines Paillier homomorphic encryption, ECDSA authentication, blockchain, and IPFS storage. The scheme supports frequent spatial aggregation for load monitoring as well as temporal aggregation for billing. However, its adversarial model explicitly assumes that the aggregator and control center do not collude. Verma et al. \cite{verma2026pphda} introduce PP-HDA, a homomorphic-signcryption-based aggregation scheme that integrates confidentiality and authentication into a single construction, thereby reducing the recurring computational and communication overhead associated with conventional encrypt-then-sign approaches.
	
	\subsection{Aggregation without Third Parties}
	
	Unlike the previous category, these approaches do not rely on trusted intermediary entities; instead, aggregation-related secrets or computations are distributed among participating entities. Such schemes can provide stronger trust decentralization, although they may introduce additional key-management, communication, or coordination overhead.
	
	Vetter et al. \cite{vetter2012homomorphic} employ additive homomorphic encryption together with a homomorphic message authentication code to support spatial and temporal aggregation. Li et al. \cite{li2010secure} construct a spanning tree over smart meters and use additive homomorphic encryption so that intermediate nodes aggregate ciphertexts without obtaining individual readings. M{\'a}rmol et al. \cite{marmol2012not} employ bihomomorphic encryption together with a dynamically selected key aggregator, preventing the energy supplier from directly decrypting individual meter contributions.
	
	Dimitriou et al. \cite{dimitriou2014secure} propose two decentralized privacy-preserving aggregation protocols: one employs symmetric cryptography and neighbor-shared randomness under semi-trusted adversaries, while the other combines public-key encryption with non-interactive zero-knowledge proofs to address active adversaries. Their subsequent work \cite{dimitriou2016secure} provides implementation and scalability evaluations. Nabil et al. \cite{nabil2019ppetd} employ pairwise secret-sharing masks that cancel during aggregation, allowing the system operator to obtain the aggregate required for load monitoring and energy management while concealing individual readings.
	
	Recent schemes further strengthen decentralized aggregation against collusion and key compromise. Wang et al. \cite{wang2026fklmpda} propose FKLM-PDA, a TTP-free multidimensional aggregation scheme that combines random masking with secret-sharing-based key separation. In addition to fault tolerance and dynamic user management, FKLM-PDA considers bounded key leakage from smart meters. Importantly, its key-leakage model is based on threshold leakage under a bounded, non-colluding setting and therefore differs from an adversarial coalition that combines exposed meter states with the views of other protocol entities.
	
	Xiao et al. \cite{xiao2026ppmmda} propose PP-MMDA, which combines ElGamal homomorphic encryption with $n$-of-$n$ distributed decryption to support multidimensional and multi-subset aggregation without relying on a trusted third party. Of particular relevance to our work, PP-MMDA explicitly considers collusion between the aggregation gateway and the control center and protects individual meter measurements against the resulting joint view. However, its adversarial model does not consider combining such infrastructure-level collusion with bounded exposure of other smart meters' internal protocol states, as considered in our work.
	
	\subsection{Anonymous Reporting via Third Parties}
	
	Another class of protocols protects customer privacy by concealing the association between fine-grained consumption values and their originating identities. In these schemes, an intermediary generally performs anonymization or manages unlinkable identities before the readings reach the utility provider.
	
	Petrlic \cite{petrlic2010privacy} combines trusted-platform mechanisms, public-key cryptography, pseudonymous credentials, and anonymization through a collector. Efthymiou et al. \cite{efthymiou2010smart} employ distinct low-frequency and high-frequency identifiers for billing and operational services, respectively. A trusted escrow manages these identifiers so that the utility provider can obtain high-frequency readings for operational purposes without directly identifying their originating customers. Although such approaches protect identity privacy, they introduce reliance on trusted anonymization entities and do not conceal the underlying fine-grained values once anonymous reports reach the authorized recipient.
	
	\subsection{Reporting via Anonymous Overlay Networks}
	
	Anonymous overlay networks provide an alternative approach in which the communication path itself is used to conceal the source of reported readings. Finster et al. \cite{finster2013pseudonymous} combine pseudonyms, blind signatures, Bloom filters, and an anonymous overlay network to transmit high-resolution readings to an untrusted grid operator. The grid operator can obtain fine-grained consumption values but cannot directly associate them with individual customers. Although this approach provides strong source anonymity, anonymous routing introduces additional communication overhead and protects the association between identity and reading rather than hiding the fine-grained reading itself from the recipient.

	Overall, previous schemes protect operational smart-meter data through
	homomorphic aggregation, secure multi-party computation, secret sharing,
	distributed masking, anonymization, and anonymous communication. More recent
	studies have strengthened privacy-preserving aggregation along different
	security and functionality dimensions, including multidimensional aggregation,
	fault tolerance, infrastructure-level collusion resistance, and smart-meter
	key-leakage resilience.
	
	In particular, Liu et al. \cite{liu2023certificateless} consider collusion
	between the control center and fog nodes, while PP-MMDA
	\cite{xiao2026ppmmda} explicitly addresses collusion between the aggregation
	gateway and the control center. In a complementary direction, FKLM-PDA
	\cite{wang2026fklmpda} considers bounded smart-meter key leakage under its
	specified leakage model. These advances address important adversarial
	capabilities individually; however, the threat model considered in this work
	targets their intersection.
	
	Specifically, we consider an adversarial coalition in which the aggregator and
	utility provider may combine their complete protocol views with the exposed
	security credentials and interval-specific internal states of a bounded number
	of smart meters, including pairwise secrets, derived keys, and generated or
	received noise components. The privacy objective is to prevent exact recovery
	of the reading of a different trusted and unexposed meter under the stated
	exposure condition, while still allowing the utility provider to obtain the
	exact aggregate required for operational services. Among the representative schemes reviewed above, prior works generally address
	infrastructure-level collusion and meter compromise as separate adversarial
	dimensions. In contrast, our threat model combines these capabilities in a
	single adversarial view: the aggregator and utility provider may collude while
	also incorporating the exposed credentials and internal protocol states of a
	bounded number of smart meters. Under this setting, the protocol aims to
	protect the exact reading of a trusted and unexposed meter while still
	providing the utility provider with the exact aggregate required for
	operational services.
	
	Table~\ref{table:related-work-comparison} provides a focused comparison of the
	privacy and threat-model dimensions most relevant to this security objective.
	
	\begin{table*}[t]
		\centering
		\caption{Comparison of representative privacy-preserving smart-metering
			schemes under relevant privacy and adversarial-model dimensions.}
		\label{table:related-work-comparison}
		
		\footnotesize
		\renewcommand{\arraystretch}{1.22}
		\setlength{\tabcolsep}{4pt}
		
		\begin{tabularx}{\textwidth}{
				>{\RaggedRight\arraybackslash}p{2.15cm}
				>{\RaggedRight\arraybackslash}X
				>{\RaggedRight\arraybackslash}p{3.35cm}
				>{\RaggedRight\arraybackslash}p{3.55cm}
				>{\RaggedRight\arraybackslash}p{2.85cm}}
			
			\toprule
			
			\textbf{Scheme} &
			\textbf{Core Privacy Mechanism} &
			\textbf{Infrastructure Collusion} &
			\textbf{Meter Exposure} &
			\textbf{Combined Model} \\
			
			\midrule
			
			Nabil et al. \cite{nabil2019ppetd} &
			Pairwise secret-sharing masks &
			-- &
			-- &
			-- \\
			
			\addlinespace[2pt]
			
			Dimitriou et al. \cite{dimitriou2014secure,dimitriou2016secure} &
			Neighbor-shared randomness / PKE with NIZK &
			-- &
			-- &
			-- \\
			
			\addlinespace[2pt]
			
			Liu et al. \cite{liu2023certificateless} &
			Certificateless cryptography and Paillier HE &
			CC--fog-node collusion &
			-- &
			-- \\
			
			\addlinespace[2pt]
			
			FKLM-PDA \cite{wang2026fklmpda} &
			Random masking and secret-sharing-based key separation &
			-- &
			Key leakage &
			-- \\
			
			\addlinespace[2pt]
			
			PP-MMDA \cite{xiao2026ppmmda} &
			ElGamal HE and distributed decryption &
			AG--CC collusion &
			-- &
			-- \\
			
			\midrule
			
			\textbf{Our scheme} &
			\textbf{Distributed noise cancellation, Paillier HE, and
				KEM--KDF--AEAD} &
			\textbf{AGG--UP collusion} &
			\textbf{Bounded credential and internal-state exposure} &
			\textbf{Explicitly considered} \\
			
			\bottomrule
			
		\end{tabularx}
		
		\vspace{1mm}
		
		\begin{minipage}{0.97\textwidth}
			\footnotesize
			\textit{Note:} ``--'' indicates that the corresponding adversarial
			capability is not explicitly modeled in the cited work. The comparison
			is limited to privacy and threat-model dimensions relevant to our
			security objective.
		\end{minipage}
	\end{table*}	

	\section{Problem Statements and Design Goals}
	As previously discussed, the transmission of fine-grained consumption data at a higher frequency can reveal customers' daily habits, presence or absence, and even results in home appliance identification, thereby compromising their privacy. On the contrary, energy suppliers require non-attributable fine-grained consumption readings for grid status analysis. To address both data privacy and utility, a collusion-resistant, privacy-preserving, aggregation-based smart metering protocol is designed to support both privacy and operational utility requirements such as real-time grid monitoring, accurate load forecasting, and maintaining demand-supply equilibrium in the smart grid network. 
	
	In the rest of this section, we first define the architecture of the smart grid along with its key components as a system model. Additionally, we outline trust levels associated with different entities within the protocol in the threat model and assumption subsections. Lastly, we explain our scheme's design goal.
	\subsection {System Model}
\begin{center}
		\begin{figure}[htp]
			\includegraphics[width=3.5in]{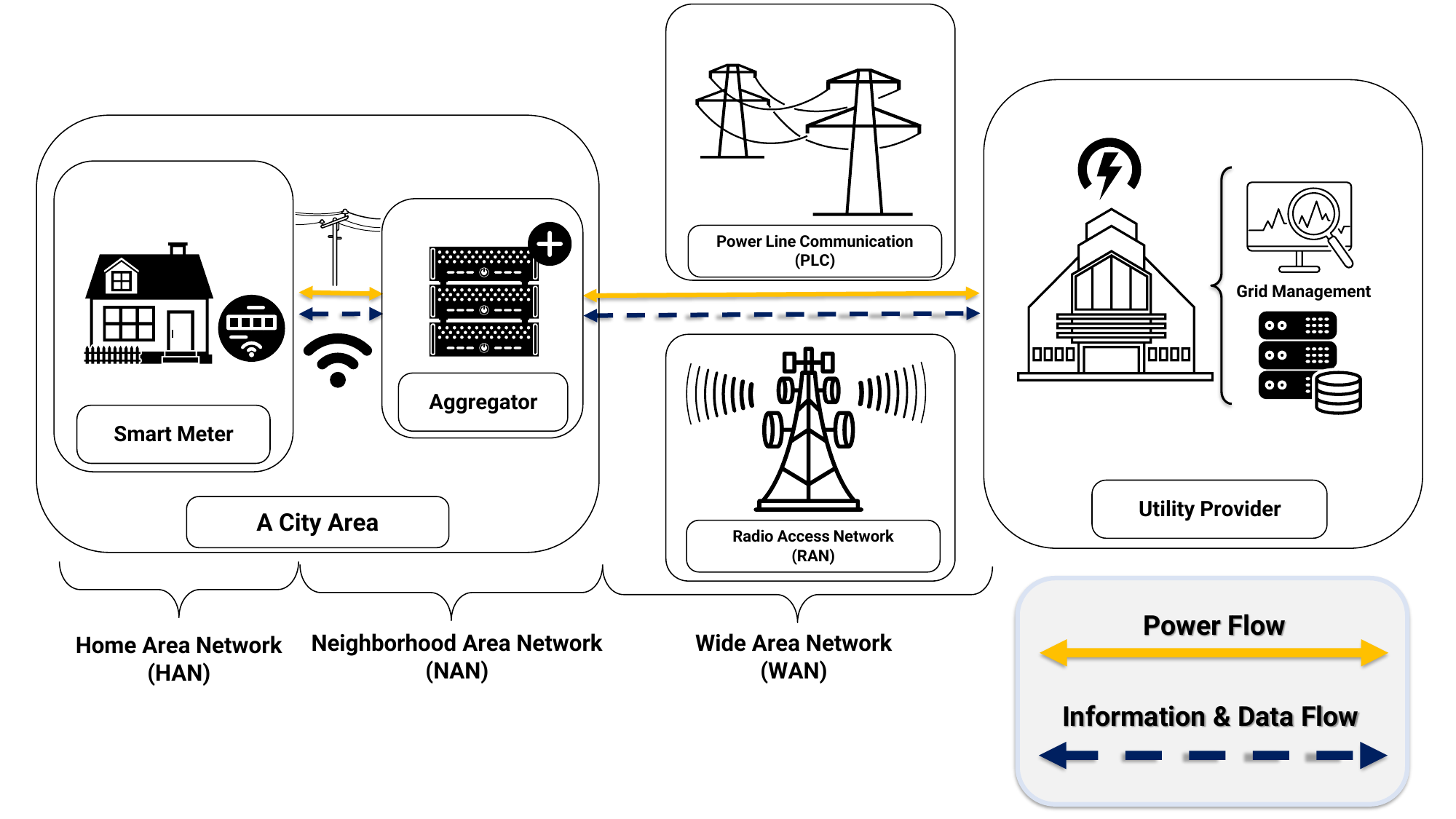}
			\caption{The smart grid's system model comprises smart meters,  aggregators, and the utility provider.}
			\label{op-system-model}
		\end{figure}
	\end{center}

	The protocol system model comprises three key entities: (1) smart meters, (2) aggregators, and (3) the utility provider. Entities collaborate and execute a specific part of the protocol to deliver a broad range of operational services within the AMI network. In the following sections, we provide a concise definition of each component within the scheme's system model. The smart grid's system model adopted in our protocol is depicted in Fig. \ref{op-system-model}.
	\begin{enumerate}
		\item \textbf{Smart Meters}: Smart meters are intelligent electronic devices that measure power usage with high precision and report it at predefined intervals (e.g., typically every 15 minutes) to an aggregator. Smart meters operate with constrained computational resources but are capable of cryptographic functions, including public-key and symmetric encryption and decryption, pairwise key establishment, key derivation, noise generation and addition, digital signing, and verification.
		\item \textbf{Aggregators}: Aggregators collect encrypted consumption contributions from smart meters, homomorphically aggregate them, and forward the encrypted aggregated result to the utility provider. They also coordinate the selection of designated smart meters and route key-establishment and authenticated-encryption messages between smart meters without obtaining the corresponding pairwise secrets or plaintext noise values. These intermediary nodes possess strong computational resources and high storage capacity. Additionally, they function as edge nodes, reducing the overall computational overhead for both smart meters and the utility provider. The energy sector specifies numerous areas for the deployment of aggregators in each city or town.
		\item \textbf{Utility Provider}: The utility provider receives Paillier-encrypted aggregated consumption values from aggregators, decrypts them using its private key, and analyzes the resulting exact aggregated values for various operational purposes, such as load forecasting, demand-supply balancing, and grid management and monitoring. The utility provider benefits from powerful computing resources and extensive storage capacity but does not obtain the pairwise secrets or plaintext noise values exchanged between smart meters.
	\end{enumerate}
	
	\subsection {Threat Model and Assumptions}
	\label{sec:threat-model-assumptions}
	As mentioned earlier, we define trust levels for each entity, including smart meters, the aggregator, and the utility provider. We also define the capabilities of an external adversary that may expose the security credentials and internal protocol states of a limited number of smart meters. We define our scheme's threat model as follows:
	\begin{itemize}
		\item \textbf{Smart Meters}: Intelligent electricity meters are fully trusted while they remain unexposed and operational, and they always adhere to the protocol. They incorporate robust tamper-resistant protection mechanisms designed to detect and respond to attempts at reverse engineering or physical intrusion, such as invasive, semi-invasive, and non-invasive hardware attacks. If tampering is detected, the meter triggers an alert to notify the control center, transitions to a non-operational state, and erases its security credentials and cached protocol secrets, including private keys, pairwise master keys, and interval-specific keys. The utility provider then investigates the incident and takes appropriate action. A detected and non-operational meter is excluded from the set of active protocol participants. The privacy guarantee is evaluated only for trusted and unexposed meters; no privacy guarantee is claimed for the individual reading or internal protocol state of a meter whose security credentials or protocol state are exposed.
		
		\item \textbf{External Adversary}: An external adversary may conduct an undetected key- or state-exposure attack against up to \(f\) smart meters. For each exposed meter, the adversary obtains its security credentials and the internal protocol state available to that meter during the reporting interval, including private keys, cached pairwise master keys, derived interval-specific keys, and generated or received noise values. However, the adversary does not modify the meter's protocol execution. Such exposure may result from temporary physical or close-range access to a meter, for example, through a maintenance or optical port, a debug or memory interface, or power-analysis and electromagnetic side-channel observations, while the meter continues to execute its legitimate firmware. For the privacy analysis, the exposed information is conservatively considered together with the complete protocol views of the aggregator and the utility provider. The set of exposed meters is determined before the designated group is selected and remains fixed during each reporting interval. Full control of a smart meter and active deviations, including false-data injection, message modification, replay, selective submission, and denial-of-service attacks, are outside the scope of this work.
				
		\item \textbf{Aggregator}: Aggregators are considered honest-but-curious edge servers. While they adhere to the established protocol, they may analyze the messages and protocol information they legitimately receive to infer additional details about customers' energy consumption. Since aggregators are configured and managed by the utility provider, collusion between these two parties is possible. Accordingly, the aggregator may share its complete protocol view with the utility provider, including the designated-group selections and the encrypted messages routed or aggregated during protocol execution. However, the aggregator does not possess the pairwise master keys or interval-specific keys established between smart meters and does not deviate from the protocol through message modification, replay, replacement, selective submission, or other active attacks.
		
		\item \textbf{Utility Provider}: Similar to the aggregator, the utility provider is considered an honest-but-curious entity and follows the established protocol. However, it may analyze the information it legitimately receives to infer additional details about consumers' power usage. The utility provider is authorized to decrypt and obtain the exact aggregated consumption value of an area at each reporting interval, but it is not authorized to obtain the exact fine-grained reading of any trusted and unexposed meter. During a collusion scenario, the utility provider and the aggregator may combine and analyze their complete protocol views. However, the utility provider does not possess the pairwise master keys or interval-specific keys established between smart meters and cannot directly decrypt the authenticated-encryption ciphertexts carrying individual noise components.
\end{itemize}

	We also established several assumptions in our protocol which are summarized as follows: 
	\begin{itemize}
		\item \textbf{Secure Communication Channels}: The communication channels between directly communicating entities in our system model are secure and authenticated, protecting messages in transit against unauthorized disclosure, modification, injection, and replay by external attackers. This assumption does not prevent an honest-but-curious endpoint or intermediary from examining messages that it legitimately receives. Therefore, meter-to-meter noise values routed through the aggregator are additionally protected using end-to-end authenticated encryption.
		
		\item \textbf{Pre-Established Entity Authentication}: Smart meters, aggregators, and the utility provider utilize an authentication scheme and register with the AMI network before protocol initiation. Only authenticated, active, and eligible smart meters are allowed to participate in pairwise key establishment, designated-meter selection, and each reporting-interval execution.

		\item \textbf{Secure Public-Key Distribution}: Public keys are securely distributed (e.g., using public-key certificates) among smart meters, aggregators, and the utility provider. This ensures that each entity obtains the authenticated public keys of the other registered entities required for Paillier encryption, signature verification, and KEM-based pairwise key establishment.

	\end{itemize}

	\subsection{Design Goals}
	Our collusion-resistant, privacy-preserving, aggregation-based smart metering protocol aims to achieve the following goals:
	\begin{enumerate}[I.]
	\item \textbf{Privacy}: The scheme aims to preserve the exact individual readings of trusted and unexposed smart meters against the adversarial view defined in Section \ref{sec:threat-model-assumptions}, provided that at least one designated and at least one non-designated smart meter remain unexposed during the reporting interval.	
	\begin{enumerate}[a)]
		\item \textbf{Preserving Fine-Grained Readings Privacy}: In our proposed protocol, individual fine-grained readings are perturbed before Paillier encryption, and only the final aggregated consumption value is decrypted by the utility provider. Consequently, the adversarial view cannot directly recover the exact reading of a trusted and unexposed smart meter under the stated exposure condition.
		
		\item \textbf{Collusion-Resistance}: The scheme aims to prevent the aggregator and the utility provider, even when colluding and combining their protocol views with the internal states of exposed smart meters, from directly recovering the exact reading of a trusted and unexposed smart meter. Under the stated exposure condition, at least one independent noise contribution relevant to the protected reading remains unknown to the adversarial coalition.
	\end{enumerate}
		\item \textbf{Data Utility}: The utilized data perturbation technique maintains data utility due to its zero-sum property. In a successful protocol execution, the applied noise values cancel during aggregation, allowing the utility provider to obtain the exact aggregated consumption value required for operational services.
		
		\item \textbf{Compatibility with Smart Grid Infrastructure}: The protocol aims to maintain computational and communication overheads compatible with the capabilities of existing smart grid devices. The hybrid construction limits each active smart meter to one Paillier encryption per reporting attempt and uses lightweight symmetric cryptography to protect the exchanged noise values. The configurable parameters \(K\) and the key-establishment epoch allow the protocol configuration to balance exposure resistance, key-management cost, and recurring protocol overhead.
	\end{enumerate}

		\section{Proposed Scheme}
		This section presents the proposed privacy-preserving aggregation protocol, first through a high-level overview and then through its detailed formal specification.
		\subsection{Protocol Overview}
		We briefly outline the steps of our proposed protocol in the following:
		
		\begin{enumerate}[i)]
			\item Before a key-establishment epoch becomes active, each pair of authenticated and eligible smart meters establishes an epoch-specific pairwise master key using an authenticated key encapsulation mechanism and a key derivation function. The aggregator only routes the required key-establishment messages and cannot obtain the established pairwise secrets.
			
			\item Before protocol execution, the configurable parameter \(K\) is selected according to the required resistance against meter exposure and the available computational and communication resources.
			
			\item At each reporting interval (e.g., every 15 minutes), the edge node, as the aggregator of a local area, uniformly selects \(K\) distinct smart meters from the set of authenticated and active meters as the designated smart meters. A new designated group is selected at each reporting interval.
			
			\item The aggregator sends a message containing the identifiers of the \(K\) designated smart meters, the reporting interval, and the protocol-attempt identifier to all active smart meters through secure and authenticated communication channels.
			
			\item Smart meters receive the message and compare the received identifiers with their own identifiers to determine whether they have been selected as designated smart meters during the current reporting interval.
			
			\item Each non-designated smart meter derives \(K\) fresh interval-specific symmetric keys from its pairwise master keys, where each derived key is associated with one of the \(K\) designated smart meters.
			
			\item Each non-designated smart meter generates \(K\) independent random values using a cryptographically secure pseudorandom number generator and a Gaussian transformation method. Each generated random value is associated with one of the \(K\) designated smart meters.
			
			\item Each non-designated smart meter computes the sum of its \(K\) generated random values and adds the resulting noise to its measured power usage to create a noisy consumption value. Since energy usage data is utilized for operational services, an appropriate data minimization technique is applied to report only the required fields (e.g., power usage in kWh).
			
			\item Each non-designated smart meter encrypts its noisy consumption value using the utility provider's Paillier public key. It also encrypts and authenticates each generated random value using the interval-specific symmetric key shared with its corresponding designated smart meter.
			
			\item Each non-designated smart meter forwards one Paillier ciphertext and \(K\) authenticated-encryption ciphertexts to the aggregator. The aggregator routes each protected random value to its corresponding designated smart meter without decrypting or aggregating it.
			
			\item Each designated smart meter decrypts and verifies the protected random values associated with it. It then computes the additive inverse of their sum and adds the resulting cancellation value to its measured energy usage. Lastly, the designated smart meter encrypts its cancellation-adjusted consumption value using the utility provider's Paillier public key and sends the resulting ciphertext to the aggregator.
			
			\item Upon receiving the Paillier-encrypted contributions from all participating smart meters, the aggregator utilizes the additive homomorphic property of the Paillier cryptosystem to compute the encrypted aggregated result. During this operation, all generated random values cancel due to their zero-sum property.
			
			\item If a required protected random value is missing or invalid, or one or more designated smart meters do not return their required contributions before a predefined timeout, the current protocol attempt is aborted. A new attempt identifier is generated, and all participating meters derive fresh interval-specific keys and generate fresh random values for the restarted attempt.
			
			\item Finally, the aggregator forwards the encrypted aggregated result to the utility provider. The utility provider decrypts and obtains the exact aggregated consumption value for various operational purposes, such as grid status analysis, management, and monitoring.
		\end{enumerate}

		In the subsequent section, we provide a detailed analysis of the proposed protocol.

		
		\subsection{Protocol Details}
		In this section, we provide the detailed formal specification of the proposed protocol. The notation used throughout the protocol is summarized in Table \ref{table:element_description_op}.
		\begin{table*}[!t]
			\centering
			\caption{Notation Used in the Proposed Protocol}
			\label{table:element_description_op}
			\scriptsize
			\renewcommand{\arraystretch}{1.12}
			\setlength{\extrarowheight}{1.2pt}
			\setlength{\tabcolsep}{4pt}
			\setlength{\arrayrulewidth}{0.3pt}
			
			\newcommand{\notationrow}[2]{%
				#1 & #2 \tabularnewline
				\cline{1-2}}
			
			\begin{tabular}{
					>{\centering\arraybackslash}p{0.23\textwidth}
					>{\raggedright\arraybackslash}p{0.71\textwidth}}
				\hline
				\rule{0pt}{2.8ex}\textbf{Symbol} & \textbf{Description} \\
				\hline
				
				\multicolumn{2}{l}{\rule{0pt}{2.7ex}\textbf{\textit{System and role notation}}} \\
				\hline
				
				\notationrow{\(D_w\)}
				{Aggregation domain or local smart-meter area.}
				
				\notationrow{\(\mathcal{M}_{D_w}\)}
				{Set of authenticated, active, and eligible smart meters participating in the current key-establishment epoch within domain \(D_w\).}
				
				\notationrow{\(M\)}
				{Number of participating smart meters, where \(M=|\mathcal{M}_{D_w}|\).}
				
				\notationrow{\(e\)}
				{Key-establishment epoch during which the pairwise master keys remain valid.}
				
				\notationrow{\(t_i\)}
				{The \(i\)th reporting interval.}
				
				\notationrow{\(a\)}
				{Protocol-attempt identifier within reporting interval \(t_i\).}
				
				\notationrow{\(K\)}
				{Number of designated smart meters selected at each reporting interval, where \(1\leq K<M\).}
				
				\notationrow{\(\mathcal{D}_{t_i}\)}
				{Set of designated smart meters at interval \(t_i\), where \(|\mathcal{D}_{t_i}|=K\).}
				
				\notationrow{\(\mathcal{N}_{t_i}\)}
				{Set of non-designated smart meters, where \(\mathcal{N}_{t_i}=\mathcal{M}_{D_w}\setminus\mathcal{D}_{t_i}\).}
				
				\notationrow{\(id_x\)}
				{Identifier of an arbitrary participating smart meter.}
				
				\notationrow{\(id_y\)}
				{Identifier of another participating smart meter that forms an unordered pair with \(id_x\) during pairwise key establishment.}
				
				\notationrow{\(id_k\)}
				{Identifier of a non-designated smart meter, where \(id_k\in\mathcal{N}_{t_i}\).}
				
				\notationrow{\(id_{sel_j}\)}
				{Identifier of the \(j\)th designated smart meter, where \(id_{sel_j}\in\mathcal{D}_{t_i}\) and \(1\leq j\leq K\).}
				
				\notationrow{\(\mathcal{E}\)}
				{Set of exposed smart meters considered in the adversarial view.}
				
				\notationrow{\(f\)}
				{Maximum number of exposed smart meters, where \(|\mathcal{E}|\leq f\).}
				
				\multicolumn{2}{l}{\rule{0pt}{2.7ex}\textbf{\textit{Reading, masking, and encoding notation}}} \\
				\hline
				
				\notationrow{\(c_{t_i,id_x}\)}
				{Actual consumption reading of meter \(id_x\) at interval \(t_i\).}
				
				\notationrow{\(C_{t_i,D_w}\)}
				{Exact aggregated consumption value of domain \(D_w\) at interval \(t_i\).}
				
				\notationrow{\(\mathcal{G}\)}
				{Cryptographically secure pseudorandom number generator used to obtain uniform samples.}
				
				\notationrow{\(\mathcal{G}\text{-}\mathrm{norm}\)}
				{Gaussian sampling procedure that combines \(\mathcal{G}\) with the adopted transformation method.}
				
				\notationrow{\(\sigma\)}
				{Target standard deviation of the total noise added by each non-designated smart meter.}
				
				\notationrow{\(\sigma_c\)}
				{Standard deviation of each independent noise component; the protocol uses \(\sigma_c=\sigma/\sqrt{K}\) to keep the total noise scale independent of \(K\).}
				
				\notationrow{\(\displaystyle s_{t_i,a,id_k}^{id_{sel_j}}\)}
				{Independent noise component generated by \(id_k\) for designated meter \(id_{sel_j}\) during attempt \(a\) of interval \(t_i\).}
				
				\notationrow{\(\displaystyle s_{t_i,a,id_k}^{total}\)}
				{Total noise generated by \(id_k\) during attempt \(a\), equal to the sum of its \(K\) independent components.}
				
				\notationrow{\(\displaystyle S_{t_i,a}^{id_{sel_j}}\)}
				{Cancellation value computed by designated meter \(id_{sel_j}\) during attempt \(a\) from the noise components assigned to it.}
				
				\notationrow{\(nc_{t_i,a,id_x}\)}
				{Noisy or cancellation-adjusted consumption value of participating meter \(id_x\) during attempt \(a\).}
				
				\notationrow{\(q\)}
				{Positive fixed-point scaling factor.}
				
				\notationrow{\(\operatorname{FP}_{q}(x)\)}
				{Signed fixed-point integer representation of a real value \(x\), defined as \(\lfloor qx\rceil\).}
				
				\notationrow{\(\operatorname{Encode}(\cdot)\)}
				{Canonical byte encoding of protocol fields prior to KDF or authenticated-encryption processing.}
				
				\notationrow{\(\displaystyle \widetilde{s}_{t_i,a,id_k}^{id_{sel_j}}\)}
				{Canonical byte encoding of \(\operatorname{FP}_{q}(s_{t_i,a,id_k}^{id_{sel_j}})\) used as the AEAD plaintext.}
				
				\notationrow{\([z]_{n_{UP}}\)}
				{Modular representation of a signed integer \(z\) in the Paillier plaintext space.}
				
				\multicolumn{2}{l}{\rule{0pt}{2.7ex}\textbf{\textit{Cryptographic notation}}} \\
				\hline
				
				\notationrow{\((pk_{UP},sk_{UP})\)}
				{Utility provider's Paillier public-private key pair.}
				
				\notationrow{\(n_{UP}\)}
				{Paillier modulus associated with \(pk_{UP}\).}
				
				\notationrow{\(\otimes\)}
				{Paillier homomorphic ciphertext-combination operation, implemented as ciphertext multiplication modulo \(n_{UP}^{2}\).}
				
				\notationrow{\(\displaystyle (pk_{id_x}^{KEM},sk_{id_x}^{KEM})\)}
				{Long-term KEM public-private key pair of meter \(id_x\), where the public key is authenticated before use.}
				
				\notationrow{\(\displaystyle ct_{x\rightarrow y,e}^{KEM}\)}
				{KEM ciphertext sent from setup initiator \(id_x\) to recipient \(id_y\) for epoch \(e\).}
				
				\notationrow{\(\displaystyle z_{\{x,y\},e}\)}
				{Raw KEM shared secret obtained by the two endpoints of unordered pair \(\{id_x,id_y\}\).}
				
				\notationrow{\(\displaystyle info_{\{x,y\},e}^{master}\)}
				{Canonical KDF context used to derive the epoch-specific pairwise master key.}
				
				\notationrow{\(\displaystyle mk_{\{x,y\},e}\)}
				{Epoch-specific pairwise master key derived for unordered pair \(\{id_x,id_y\}\).}
				
				\notationrow{\(\displaystyle ak_{t_i,a,id_k}^{id_{sel_j}}\)}
				{Directional interval-specific AEAD key derived for transmission from \(id_k\) to \(id_{sel_j}\).}
				
				\notationrow{\(\nu_{t_i,a,k,j}\)}
				{AEAD nonce associated with the corresponding noise-component record.}
				
				\notationrow{\(AAD_{t_i,a,k,j}\)}
				{Associated data that binds an AEAD ciphertext to its domain, epoch, interval, attempt, sender, recipient, and message type.}
				
				\notationrow{\(\displaystyle \mathcal{C}_{t_i,a,id_{sel_j}}^{k}\)}
				{AEAD ciphertext protecting \(\widetilde{s}_{t_i,a,id_k}^{id_{sel_j}}\) during attempt \(a\).}
				
				\notationrow{\(\mathcal{X}_{UP,t_i,a}^{x}\)}
				{Paillier ciphertext of meter \(id_x\)'s encoded noisy or cancellation-adjusted consumption value during attempt \(a\).}
				
				\notationrow{\(\mathscr{X}_{t_i,a,D_w}\)}
				{Final Paillier ciphertext containing the encrypted aggregate for attempt \(a\) of interval \(t_i\) in domain \(D_w\).}
				
				\hline
			\end{tabular}
		\end{table*}
		As we mentioned earlier in Section \ref{sec:threat-model-assumptions}, the protocol considers three key assumptions: secure and authenticated communication channels between directly communicating entities, authentication of entities before participating in the protocol, and secure distribution of authenticated public keys before protocol execution. According to these assumptions, each entity obtains the authenticated public
		keys required for Paillier encryption and KEM-based pairwise key
		establishment. The communication channels protect protocol messages in transit against unauthorized disclosure, modification, injection, and replay by external attackers. However, since the aggregator may legitimately receive and route messages exchanged between smart meters, the individual noise components are additionally protected using end-to-end authenticated encryption.
		
		\par\noindent\textbf{Phase I -- Pairwise KEM--KDF Master-Key Establishment:}
		This phase establishes one epoch-specific master key between every unordered
		pair of authenticated and eligible smart meters. The established master keys
		remain valid during epoch \(e\) and are subsequently used to derive fresh
		interval-specific keys without repeating the KEM operations during every
		reporting interval.
		
		Before a key-establishment epoch \(e\) becomes active, every unordered pair of authenticated and eligible smart meters in \(\mathcal{M}_{D_w}\) establishes one epoch-specific pairwise master key. For each unordered pair \(\{id_x,id_y\}\), a deterministic function \(F\) selects one endpoint as the setup initiator. This selection can distribute the encapsulation operations among the smart meters and avoid assigning the complete setup workload to a single meter.
		\begin{IEEEeqnarray}{rCl}
			I_{x,y,e} &\leftarrow& F(D_w,e,id_x,id_y), \qquad
			I_{x,y,e}\in\{id_x,id_y\}.
			\label{eq:op-protocol-details-kem-initiator}
		\end{IEEEeqnarray}
		Let \(id_I=I_{x,y,e}\) denote the selected initiator and let \(id_R\) denote the other endpoint of the pair. The initiator performs the KEM encapsulation using the authenticated KEM public key of \(id_R\). This operation generates a KEM ciphertext and a raw shared secret.
		\begin{IEEEeqnarray}{rCl}
			\left(ct_{I\rightarrow R,e}^{KEM},
			z_{\{x,y\},e}^{(I)}\right)
			&\leftarrow&
			KEM.Encaps\left(pk_{id_R}^{KEM}\right).
			\label{eq:op-protocol-details-kem-encapsulation}
		\end{IEEEeqnarray}
		The initiator sends the KEM ciphertext, together with the corresponding domain, epoch, initiator, and recipient identifiers, to the aggregator. The aggregator only routes this setup message to \(id_R\) and does not obtain the raw shared secret.
		\begin{IEEEeqnarray}{rCl}
			SM_I &\xrightarrow{}& AGG:
			D_w\Vert e\Vert id_I\Vert id_R\Vert
			ct_{I\rightarrow R,e}^{KEM},
			\notag\\
			AGG &\xrightarrow{}& SM_R:
			D_w\Vert e\Vert id_I\Vert id_R\Vert
			ct_{I\rightarrow R,e}^{KEM}.
			\label{eq:op-protocol-details-kem-routing}
		\end{IEEEeqnarray}
		After receiving and validating the routed setup message, \(id_R\) decapsulates the KEM ciphertext using its KEM private key.
		\begin{IEEEeqnarray}{rCl}
			z_{\{x,y\},e}^{(R)}
			&\leftarrow&
			KEM.Decaps\left(
			sk_{id_R}^{KEM},
			ct_{I\rightarrow R,e}^{KEM}
			\right).
			\label{eq:op-protocol-details-kem-decapsulation}
		\end{IEEEeqnarray}
		For a valid KEM execution, both endpoints obtain the same raw shared secret:
		\begin{IEEEeqnarray}{rCl}
			z_{\{x,y\},e}^{(I)}
			&=&
			z_{\{x,y\},e}^{(R)}
			=
			z_{\{x,y\},e}.
			\label{eq:op-protocol-details-kem-correctness}
		\end{IEEEeqnarray}
		The raw KEM shared secret is not directly used as an authenticated-encryption key. Instead, both endpoints first construct the same canonical master-key context, where \(id_{\min}\) and \(id_{\max}\) denote the identifiers of the pair in canonical order.
		\begin{IEEEeqnarray}{rCl}
			info_{\{x,y\},e}^{master}
			&\leftarrow&
			\operatorname{Encode}\left(
			\begin{aligned}
				&D_w
				\Vert e
				\Vert id_{\min}
				\Vert id_{\max} \\
				&\Vert \texttt{pairwise-master}
			\end{aligned}
			\right).
			\label{eq:op-protocol-details-master-key-context}
		\end{IEEEeqnarray}
		Both smart meters then derive the epoch-specific pairwise master key as follows:
		\begin{IEEEeqnarray}{rCl}
			mk_{\{x,y\},e}
			&\leftarrow&
			KDF\left(
			z_{\{x,y\},e},
			info_{\{x,y\},e}^{master}
			\right).
			\label{eq:op-protocol-details-master-key-derivation}
		\end{IEEEeqnarray}
		After successful master-key derivation, both endpoints securely erase the raw KEM shared secret and any temporary encapsulation or decapsulation state.
		Since one pairwise master key is established for every unordered pair of
		smart meters in \(\mathcal{M}_{D_w}\), where
		\(M=|\mathcal{M}_{D_w}|\) denotes the number of participating smart meters,
		the total number of distinct pairwise master keys in epoch \(e\) is:
		
		\begin{IEEEeqnarray}{rCl}
			\left|
			\left\{
			mk_{\{x,y\},e}
			\,\middle|\,
			1\leq x<y\leq M
			\right\}
			\right|
			&=&
			\binom{M}{2}
			=
			\frac{M(M-1)}{2}.
			\label{eq:op-protocol-details-number-pairwise-master-keys}
		\end{IEEEeqnarray}
		Each pairwise master key is securely stored by its two corresponding endpoints. Consequently, every smart meter stores \(M-1\) pairwise master keys during the active epoch, while the total number of stored key copies across the aggregation domain is:
		\begin{IEEEeqnarray}{rCl}
			N_{mk,id_x} &=& M-1,
			\qquad id_x\in\mathcal{M}_{D_w},
			\notag\\
			N_{mk}^{copies}
			&=&
			2\binom{M}{2}
			=
			M(M-1).
			\label{eq:op-protocol-details-master-key-storage}
		\end{IEEEeqnarray}
		The key-establishment epoch becomes active only after all required meter pairs have successfully established their corresponding master keys. A smart meter that has not completed the required pairwise setup is excluded from the set of eligible protocol participants. The established master keys remain valid only during epoch \(e\) and are securely erased when the epoch expires. Therefore, the KEM operations are completed before the reporting intervals begin, while the active protocol execution uses the stored pairwise master keys to derive fresh interval-specific keys.
		
		\par\noindent\textbf{Phase II -- Designated-Meter Selection and Role Assignment:}
		This phase selects the \(K\) designated smart meters for reporting interval
		\(t_i\), distributes the authenticated interval and attempt context, and
		enables every participating smart meter to determine its role in the current
		protocol execution.
		
		At each reporting interval (e.g., every 15 minutes), the local aggregator uniformly selects \(K\) distinct smart meters from the set of authenticated, active, and eligible meters \(\mathcal{M}_{D_w}\) as the designated smart meters. The remaining participating meters are considered non-designated smart meters. Therefore, the designated and non-designated meter sets for interval \(t_i\) are defined as follows:
		\begin{IEEEeqnarray}{rCl}
			\mathcal{D}_{t_i}
			&\xleftarrow{\$}&
			\left\{
			\mathcal{D}\subseteq\mathcal{M}_{D_w}
			\,\middle|\,
			|\mathcal{D}|=K
			\right\},
			\notag\\
			\mathcal{D}_{t_i}
			&=&
			\left\{
			id_{sel_1},id_{sel_2},\ldots,id_{sel_K}
			\right\},
			\notag\\
			\mathcal{N}_{t_i}
			&=&
			\mathcal{M}_{D_w}\setminus\mathcal{D}_{t_i},
			\qquad
			|\mathcal{N}_{t_i}|=M-K.
			\label{eq:op-protocol-details-designated-set-selection}
		\end{IEEEeqnarray}
		A new designated group is selected at each reporting interval to distribute the noise-cancellation responsibility among the participating smart meters. After selecting the designated group, the aggregator forwards an authenticated interval message containing the aggregation domain (\(D_w\)), the active key-establishment epoch (\(e\)),
		the reporting interval (\(t_i\)), the protocol-attempt identifier (\(a\)),
		and the designated-meter set (\(\mathcal{D}_{t_i}\)), containing the
		identifiers of the \(K\) designated smart meters, to all participating meters
		through secure and authenticated communication channels.
		
		\begin{IEEEeqnarray}{rCl}
			AGG
			&\xrightarrow{}&
			\mathrm{SMs}:
			D_w
			\Vert e
			\Vert t_i
			\Vert a
			\Vert \mathcal{D}_{t_i}.
			\label{eq:op-protocol-details-designated-set-message}
		\end{IEEEeqnarray}
		Afterward, each smart meter receives the interval message and compares its own identifier with the identifiers contained in \(\mathcal{D}_{t_i}\). This allows each meter to determine whether it has been selected as a designated smart meter during the current reporting interval. 
		If the received interval message is inconsistent with the active protocol context, the smart meter rejects the message, and the corresponding protocol attempt is aborted according to the protocol's failure-handling policy.
		
		\par\noindent\textbf{Phase III -- Noise-Component Generation and Noisy-Reading Construction:}
		This phase requires each non-designated smart meter to generate one
		independent Gaussian noise component for every designated smart meter. The
		component variance is selected so that their sum has the target noise
		standard deviation \(\sigma\), independently of the configured value of
		\(K\).
		
		After determining their roles, the non-designated smart meters employ the noise addition algorithm \(\mathcal{G}\text{-}\mathrm{norm}\), which combines a cryptographically secure pseudorandom number generator \(\mathcal{G}\) with a Gaussian transformation method. Each non-designated smart meter \(id_k\in\mathcal{N}_{t_i}\) generates one independent noise component for every designated smart meter \(id_{sel_j}\in\mathcal{D}_{t_i}\). To obtain a total noise standard deviation of \(\sigma\) independently of the number of designated smart meters, the standard deviation of each noise component is defined as:
		\begin{IEEEeqnarray}{rCl}
			\sigma_c
			&=&
			\frac{\sigma}{\sqrt{K}},
			\qquad
			K\sigma_c^{2}=\sigma^{2}.
			\label{eq:op-protocol-details-component-noise-standard-deviation}
		\end{IEEEeqnarray}
		Accordingly, for every \(id_k\in\mathcal{N}_{t_i}\) and \(j\in\{1,\ldots,K\}\), the meter generates:
		\begin{IEEEeqnarray}{rCl}
			s_{t_i,a,id_k}^{id_{sel_j}}
			&\leftarrow&
			\mathcal{G}\text{-}\mathrm{norm}
			\left(\mathcal{G},\sigma_c\right),
			\qquad
			s_{t_i,a,id_k}^{id_{sel_j}}
			\sim
			\mathcal{N}\left(0,\sigma_c^{2}\right).
			\label{eq:op-protocol-details-noise-component-generation}
		\end{IEEEeqnarray}
		The generated noise components are fresh and mutually independent across different non-designated meters, designated recipients, reporting intervals, and protocol attempts. Therefore, if an attempt is aborted and restarted, every non-designated smart meter discards its previously generated noise components and generates fresh components for the new attempt.

		For a fixed non-designated smart meter \(id_k\), reporting interval \(t_i\), and protocol attempt \(a\), the Gaussian sampling procedure is implemented using the Box--Muller transformation. Since each execution of the Box--Muller transformation generates two Gaussian samples, the meter executes the transformation
		\(\lceil K/2\rceil\) times to obtain at least \(K\) noise components. 
		Each execution requires two independent uniform values. Therefore, using the
		fresh private state of its cryptographically secure pseudorandom number
		generator (CSPRNG) \(\mathcal{G}\), the meter generates
		\(2\lceil K/2\rceil\) pseudorandom \(L\)-bit strings. Let \(r_{\ell}\)
		denote the \(\ell\)-th generated string, where
		\(\ell\in\{1,\ldots,2\lceil K/2\rceil\}\), and let
		\(\operatorname{int}(r_{\ell})\) denote its integer representation. The
		corresponding uniform values are obtained as:
		\begin{IEEEeqnarray}{rCl}
			r_{\ell}
			&\leftarrow&
			\mathcal{G},
			\notag\\
			u_{\ell}
			&=&
			\frac{\operatorname{int}(r_{\ell})+1}{2^{L}+1},
			\qquad
			\ell\in\left\{1,\ldots,2\left\lceil\frac{K}{2}\right\rceil\right\}.
			\label{eq:op-protocol-details-uniform-sample-generation}
		\end{IEEEeqnarray}
		
		For each \(h\in\{1,\ldots,\lceil K/2\rceil\}\), two Gaussian samples with
		standard deviation \(\sigma_c\) are then generated from \(u_{2h-1}\) and
		\(u_{2h}\):
		```
		
		\begin{IEEEeqnarray}{rCl}
			R_h
			&=&
			\sqrt{-2\ln\left(u_{2h-1}\right)},
			\notag\\
			\Theta_h
			&=&
			2\pi u_{2h},
			\notag\\
			B_{2h-1}
			&=&
			\sigma_c R_h\cos\left(\Theta_h\right),
			\notag\\
			B_{2h}
			&=&
			\sigma_c R_h\sin\left(\Theta_h\right).
			\label{eq:op-protocol-details-box-muller-transformation}
		\end{IEEEeqnarray}
		The first \(K\) generated samples are sequentially assigned to the \(K\) designated smart meters:
		\begin{IEEEeqnarray}{rCl}
			s_{t_i,a,id_k}^{id_{sel_j}}
			&=&
			B_j,
			\qquad
			j=1,\ldots,K.
			\label{eq:op-protocol-details-noise-component-assignment}
		\end{IEEEeqnarray}
		If \(K\) is odd, only the final unused sample \(B_{K+1}\) is discarded. 
		
		After generating the \(K\) noise components, each non-designated smart meter \(id_k\in\mathcal{N}_{t_i}\) computes its total noise by summing the components assigned to the designated smart meters:
		\begin{IEEEeqnarray}{rCl}
			s_{t_i,a,id_k}^{total}
			&=&
			\sum_{j=1}^{K}
			s_{t_i,a,id_k}^{id_{sel_j}}.
			\label{eq:op-protocol-details-total-noise-computation}
		\end{IEEEeqnarray}
		Since the \(K\) components are mutually independent and each follows
		\(\mathcal{N}(0,\sigma_c^2)\), their sum follows a Gaussian distribution with variance \(K\sigma_c^2=\sigma^2\):
		\begin{IEEEeqnarray}{rCl}
			s_{t_i,a,id_k}^{total}
			&\sim&
			\mathcal{N}\left(
			0,
			K\sigma_c^2
			\right)
			=
			\mathcal{N}\left(0,\sigma^2\right).
			\label{eq:op-protocol-details-total-noise-distribution}
		\end{IEEEeqnarray}
		The non-designated smart meter then adds the total noise to its actual consumption reading \(c_{t_i,id_k}\) to obtain its noisy consumption value:
		\begin{IEEEeqnarray}{rCl}
			nc_{t_i,a,id_k}
			&=&
			c_{t_i,id_k}
			+
			s_{t_i,a,id_k}^{total}.
			\label{eq:op-protocol-details-non-designated-noisy-consumption}
		\end{IEEEeqnarray}
		
		\par\noindent\textbf{Phase IV -- Interval-Key Derivation and AEAD Protection:}
		This phase derives a directional, attempt-specific AEAD key for every
		non-designated-to-designated meter transmission. Each noise component is
		converted once to its fixed-point integer representation and is then
		protected using AEAD with context-binding associated data and a fresh nonce.
		
		After computing its noisy consumption value, each non-designated smart meter \(id_k\in\mathcal{N}_{t_i}\) derives one directional interval-specific AEAD key for every designated smart meter \(id_{sel_j}\in\mathcal{D}_{t_i}\). Both endpoints first construct the same interval-key context:
		\begin{IEEEeqnarray}{rCl}
			info_{t_i,a,k,j}^{interval}
			&\leftarrow&
			\operatorname{Encode}\left(
			\begin{aligned}
				&D_w
				\Vert e
				\Vert t_i
				\Vert a 
				\Vert id_k
				\Vert id_{sel_j}\\
				&\Vert \texttt{mask-encryption}
			\end{aligned}
			\right).
			\label{eq:op-protocol-details-interval-key-context}
		\end{IEEEeqnarray}
		Using their previously established pairwise master key, both endpoints independently derive:
		\begin{IEEEeqnarray}{rCl}
			ak_{t_i,a,id_k}^{id_{sel_j}}
			&\leftarrow&
			KDF\left(
			mk_{\{k,sel_j\},e},
			info_{t_i,a,k,j}^{interval}
			\right).
			\label{eq:op-protocol-details-interval-key-derivation}
		\end{IEEEeqnarray}
		The sender and recipient identifiers associate the derived key with the transmission from \(id_k\) to \(id_{sel_j}\). Moreover, including \(t_i\) and \(a\) in the KDF context causes the same meter pair to derive a different interval-specific key for each reporting interval and protocol attempt.
		
		Before encrypting a noise component, the non-designated smart meter \(id_k\) constructs the associated data for the corresponding transmission to \(id_{sel_j}\). The associated data identifies the domain, epoch, reporting interval, protocol attempt, sender, recipient, and message type:
		\begin{IEEEeqnarray}{rCl}
			AAD_{t_i,a,k,j}
			&\leftarrow&
			\operatorname{Encode}\left(
			\begin{aligned}
				&D_w
				\Vert e
				\Vert t_i
				\Vert a 
				\Vert id_k
				\Vert id_{sel_j}\\
				&\Vert \texttt{mask-record}
			\end{aligned}
			\right).
			\label{eq:op-protocol-details-aead-associated-data}
		\end{IEEEeqnarray}
		The associated data is not encrypted, but it is authenticated together with the noise component. Therefore, the designated smart meter must use the same associated data when verifying and decrypting the received AEAD ciphertext.
				
		For each noise-component record sent from \(id_k\) to \(id_{sel_j}\), the non-designated smart meter generates a nonce of the length required by the selected AEAD algorithm:
		\begin{IEEEeqnarray}{rCl}
			\nu_{t_i,a,k,j}
			&\xleftarrow{\$}&
			\{0,1\}^{\ell_{\nu}},
			\label{eq:op-protocol-details-aead-nonce-generation}
		\end{IEEEeqnarray}
		where \(\ell_{\nu}\) denotes the required AEAD nonce length. The nonce must not be reused with the same interval-specific key. Therefore, each interval-specific key and its corresponding nonce are used for exactly one AEAD encryption operation. The non-designated smart meter retains the generated AEAD ciphertext only until the corresponding protocol attempt is successfully completed or aborted. During this period, the same ciphertext may be retransmitted when necessary. After completion or abortion of the attempt, the meter securely erases the ciphertext together with the corresponding interval-specific key, nonce, and noise component.
		
		Before encrypting a noise component, the non-designated smart meter converts it once to a signed fixed-point integer and canonically encodes the resulting value:
		\begin{IEEEeqnarray}{rCl}
			\widetilde{s}_{t_i,a,id_k}^{id_{sel_j}}
			&\leftarrow&
			\operatorname{Encode}\left(
			\operatorname{FP}_{q}\left(
			s_{t_i,a,id_k}^{id_{sel_j}}
			\right)
			\right).
			\label{eq:op-protocol-details-noise-component-encoding}
		\end{IEEEeqnarray}
		The fixed-point integer is retained throughout the current protocol attempt. The same integer representation is used when forming the encoded noisy consumption value and when the designated smart meter later computes the corresponding cancellation value. This prevents independent rounding operations from causing incomplete noise cancellation. The non-designated smart meter then encrypts the encoded noise component using the corresponding interval-specific key, nonce, and associated data:
		\begin{IEEEeqnarray}{rCl}
			\mathcal{C}_{t_i,a,id_{sel_j}}^{k}
			&\leftarrow&
			AEAD.Enc\left(
			\begin{aligned}
				&ak_{t_i,a,id_k}^{id_{sel_j}},
				\ \nu_{t_i,a,k,j},\\
				&\widetilde{s}_{t_i,a,id_k}^{id_{sel_j}},
				\ AAD_{t_i,a,k,j}
			\end{aligned}
			\right).
			\label{eq:op-protocol-details-noise-component-aead-encryption}
		\end{IEEEeqnarray}
		
		\par\noindent\textbf{Phase V -- Non-Designated Contribution Preparation and Submission:}
		This phase constructs the Paillier contribution of each non-designated smart
		meter using the same fixed-point noise integers carried in its AEAD records.
		The meter then packages its Paillier ciphertext and recipient-specific
		noise-component records and submits them to the aggregator.
		
		The non-designated smart meter next prepares its noisy consumption value for Paillier encryption. Since Paillier encrypts integers, the meter performs this operation in the fixed-point domain. It first sums the fixed-point integers corresponding to its \(K\) noise components:
		\begin{IEEEeqnarray}{rCl}
			z_{t_i,a,id_k}^{mask}
			&=&
			\sum_{j=1}^{K}
			\operatorname{FP}_{q}\left(
			s_{t_i,a,id_k}^{id_{sel_j}}
			\right).
			\notag
		\end{IEEEeqnarray}
		It then adds this integer mask sum to the fixed-point representation of its consumption reading and maps the resulting signed integer to the Paillier plaintext space:
		\begin{IEEEeqnarray}{rCl}
			m_{t_i,a,id_k}^{ND}
			&=&
			\left[
			\operatorname{FP}_{q}\left(c_{t_i,id_k}\right)
			+
			z_{t_i,a,id_k}^{mask}
			\right]_{n_{UP}}.
			\notag
		\end{IEEEeqnarray}
		Finally, the meter encrypts the resulting plaintext using the utility provider's Paillier public key:
		\begin{IEEEeqnarray}{rCl}
			\mathcal{X}_{UP,t_i,a}^{k}
			&\leftarrow&
			\operatorname{Paillier.Enc}\left(
			pk_{UP},
			m_{t_i,a,id_k}^{ND}
			\right).
			\label{eq:op-protocol-details-non-designated-paillier-encryption}
		\end{IEEEeqnarray}
		The fixed-point noise integers used in \(z_{t_i,a,id_k}^{mask}\) are the same integers previously encoded in the corresponding AEAD plaintexts; they are not computed again from the real-valued noise samples. This ensures that the designated smart meters later subtract exactly the same integer values. Therefore, each non-designated smart meter performs one Paillier encryption during each protocol attempt, independently of \(K\).
		
		After generating its Paillier ciphertext and the \(K\) AEAD ciphertexts, each non-designated smart meter \(id_k\in\mathcal{N}_{t_i}\) prepares one noise-component record for every designated smart meter. Each record contains the intended recipient identifier, the corresponding AEAD nonce, and the AEAD ciphertext:
		\begin{IEEEeqnarray}{rCl}
			\mathcal{R}_{t_i,a,k,j}
			&=&
			id_{sel_j}
			\Vert
			\nu_{t_i,a,k,j}
			\Vert
			\mathcal{C}_{t_i,a,id_{sel_j}}^{k},
			\qquad
			j\in[1..K].
			\label{eq:op-protocol-details-noise-component-record}
		\end{IEEEeqnarray}
		The non-designated smart meter then sends its Paillier ciphertext and the \(K\) noise-component records to the aggregator together with the protocol-context identifiers:
		\begin{IEEEeqnarray}{rCl}
			SM_k
			&\xrightarrow{}&
			AGG:
			D_w
			\Vert e
			\Vert t_i
			\Vert a
			\Vert id_k
			\notag\\
			&&\quad
			\Vert\mathcal{X}_{UP,t_i,a}^{k}
			\Vert
			\mathcal{R}_{t_i,a,k,1}
			\Vert\cdots\Vert
			\mathcal{R}_{t_i,a,k,K}.
			\label{eq:op-protocol-details-non-designated-transmission}
		\end{IEEEeqnarray}
		The protocol-context identifiers allow each noise-component record to be associated with its reporting interval, protocol attempt, sender, and intended designated recipient.
		
		\par\noindent\textbf{Phase VI -- Secure Routing and Designated-Meter Cancellation:}
		This phase routes each AEAD-protected noise component to its designated
		recipient while retaining the corresponding Paillier contribution for final
		aggregation. After successfully verifying all expected records, each
		designated smart meter computes the exact fixed-point cancellation value and
		encrypts its cancellation-adjusted consumption contribution.
		
		Upon receiving the contribution of a non-designated smart meter \(id_k\), the aggregator separates the Paillier ciphertext from the \(K\) noise-component records. It stores the Paillier ciphertext for the final homomorphic aggregation:
		\begin{IEEEeqnarray}{rCl}
			AGG
			&:&
			\operatorname{Store}\left(
			\mathcal{X}_{UP,t_i,a}^{k}
			\right).
			\label{eq:op-protocol-details-aggregator-store-paillier}
		\end{IEEEeqnarray}
		For each \(j\in\{1,\ldots,K\}\), the aggregator forwards the corresponding noise-component record to its designated recipient:
		\begin{IEEEeqnarray}{rCl}
			AGG
			&\xrightarrow{}&
			SM_{sel_j}:
			D_w
			\Vert e
			\Vert t_i
			\Vert a
			\Vert id_k
			\notag\\
			&&\quad
			\Vert\mathcal{R}_{t_i,a,k,j}.
			\label{eq:op-protocol-details-aggregator-route-noise-record}
		\end{IEEEeqnarray}
		The aggregator only routes the AEAD-protected noise-component records. It neither decrypts these records nor combines their ciphertexts, since AEAD ciphertexts do not support the Paillier homomorphic aggregation operation.
		
		Each designated smart meter \(id_{sel_j}\in\mathcal{D}_{t_i}\) receives one noise-component record from every non-designated smart meter \(id_k\in\mathcal{N}_{t_i}\). Using the sender identifier and the protocol-context fields, the designated meter reconstructs the corresponding interval-specific key and associated data. It then verifies and decrypts each AEAD ciphertext as follows:
		\begin{IEEEeqnarray}{rCl}
			\widetilde{s}_{t_i,a,id_k}^{id_{sel_j}}
			&\leftarrow&
			\operatorname{AEAD.Dec}\left(
			\begin{array}{l}
				ak_{t_i,a,id_k}^{id_{sel_j}},\\
				\nu_{t_i,a,k,j},\\
				\mathcal{C}_{t_i,a,id_{sel_j}}^{k},\\
				AAD_{t_i,a,k,j}
			\end{array}
			\right).
			\label{eq:op-protocol-details-noise-component-aead-decryption}
		\end{IEEEeqnarray}
		If any expected record is missing or any AEAD decryption returns \(\bot\), the entire protocol attempt is aborted according to the failure-handling policy. The designated smart meter computes its cancellation contribution only after all \(M-K\) expected noise-component records have been successfully received and verified.
		
		After all expected AEAD records have been successfully verified and decrypted, each designated smart meter \(id_{sel_j}\) obtains the noise components sent to it by the \(M-K\) non-designated smart meters. It computes its cancellation value as the negative sum of these components and adds the result to its own consumption reading:
		\begin{IEEEeqnarray}{rCl}
			S_{t_i,a}^{id_{sel_j}}
			&=&
			-\sum_{id_k\in\mathcal{N}_{t_i}}
			s_{t_i,a,id_k}^{id_{sel_j}},
			\notag\\
			nc_{t_i,a,id_{sel_j}}
			&=&
			c_{t_i,id_{sel_j}}
			+
			S_{t_i,a}^{id_{sel_j}}.
			\label{eq:op-protocol-details-designated-cancellation}
		\end{IEEEeqnarray}
		In the actual computation, the designated smart meter performs this cancellation using the fixed-point integers recovered from the AEAD plaintexts. It does not independently re-encode the real-valued noise components, which ensures that the same integer values previously added by the non-designated smart meters are subtracted exactly.
		
		The designated smart meter next prepares its cancellation-adjusted contribution for Paillier encryption. The value \(nc_{t_i,a,id_{sel_j}}\) represents the corresponding real-valued quantity, whereas Paillier requires an integer plaintext. Therefore, the meter constructs the plaintext directly in the fixed-point domain:
		\begin{IEEEeqnarray}{rCl}
			m_{t_i,a,id_{sel_j}}^{D}
			&=&
			\left[
			\begin{array}{l}
				\operatorname{FP}_{q}\left(c_{t_i,id_{sel_j}}\right)
				\\[0.3ex]
				\displaystyle {}-
				\sum_{id_k\in\mathcal{N}_{t_i}}
				\operatorname{FP}_{q}\left(
				s_{t_i,a,id_k}^{id_{sel_j}}
				\right)
			\end{array}
			\right]_{n_{UP}}.
			\label{eq:op-protocol-details-designated-plaintext}
		\end{IEEEeqnarray}
		The fixed-point noise integers in this sum are exactly those recovered from the verified AEAD plaintexts. The designated smart meter then encrypts the resulting plaintext using the utility provider's Paillier public key:
		\begin{IEEEeqnarray}{rCl}
			\mathcal{X}_{UP,t_i,a}^{sel_j}
			&\leftarrow&
			\operatorname{Paillier.Enc}\left(
			pk_{UP},
			m_{t_i,a,id_{sel_j}}^{D}
			\right).
			\label{eq:op-protocol-details-designated-paillier-encryption}
		\end{IEEEeqnarray}
		Thus, each designated smart meter produces one Paillier ciphertext for the current protocol attempt and sends it to the aggregator.
		
		\par\noindent\textbf{Phase VII -- Homomorphic Aggregation and Aggregate Recovery:}
		This phase combines exactly one Paillier ciphertext from every participating
		smart meter into a single encrypted aggregate. The utility provider then
		decrypts and decodes the aggregate ciphertext to recover the fixed-point
		aggregate consumption value for reporting interval \(t_i\).
		
		After receiving exactly one Paillier ciphertext from every participating
		smart meter for the same protocol attempt \((D_w,e,t_i,a)\), the aggregator
		combines the \(M\) ciphertexts using the Paillier homomorphic operation:
		\begin{IEEEeqnarray}{rCl}
			\mathscr{X}_{t_i,a,D_w}
			&=&
			\bigotimes_{id_x\in\mathcal{M}_{D_w}}
			\mathcal{X}_{UP,t_i,a}^{x}
			\notag\\
			&=&
			\left(
			\prod_{id_x\in\mathcal{M}_{D_w}}
			\mathcal{X}_{UP,t_i,a}^{x}
			\right)
			\bmod n_{UP}^{2}.
			\label{eq:op-protocol-details-final-paillier-aggregation}
		\end{IEEEeqnarray}
		The resulting ciphertext \(\mathscr{X}_{t_i,a,D_w}\) encrypts the sum of the
		fixed-point plaintext contributions of all participating smart meters. The
		aggregator does not decrypt any individual contribution and forwards only the
		final aggregate ciphertext to the utility provider:
		\begin{IEEEeqnarray}{rCl}
			AGG
			&\xrightarrow{}&
			UP:
			D_w
			\Vert e
			\Vert t_i
			\Vert a
			\Vert
			\mathscr{X}_{t_i,a,D_w}.
			\label{eq:op-protocol-details-final-ciphertext-transmission}
		\end{IEEEeqnarray}
		If any participating meter's Paillier ciphertext is missing when the attempt
		timeout expires, the aggregator does not compute the final aggregate, and the
		corresponding protocol attempt is aborted.
		
		Upon receiving the final aggregate ciphertext, the utility provider decrypts it using its Paillier private key:
		\begin{IEEEeqnarray}{rCl}
			m_{t_i,a,D_w}^{agg}
			&\leftarrow&
			\operatorname{Paillier.Dec}\left(
			sk_{UP},
			\mathscr{X}_{t_i,a,D_w}
			\right).
			\label{eq:op-protocol-details-final-paillier-decryption}
		\end{IEEEeqnarray}
		To prevent modular wraparound, the protocol parameters must ensure that the sum of the fixed-point consumption readings satisfies:
		\begin{IEEEeqnarray}{rCl}
			0
			&\leq&
			\sum_{id_x\in\mathcal{M}_{D_w}}
			\operatorname{FP}_{q}\left(c_{t_i,id_x}\right)
			<n_{UP}.
			\label{eq:op-protocol-details-aggregate-no-wraparound}
		\end{IEEEeqnarray}
		Under this condition, \(m_{t_i,a,D_w}^{agg}\) is the integer aggregate of the participating meters' fixed-point consumption readings. The utility provider recovers the aggregate consumption value by applying the inverse scaling factor:
		\begin{IEEEeqnarray}{rCl}
			C_{t_i,D_w}
			&=&
			\frac{m_{t_i,a,D_w}^{agg}}{q}
			\notag\\
			&=&
			\frac{1}{q}
			\sum_{id_x\in\mathcal{M}_{D_w}}
			\operatorname{FP}_{q}\left(c_{t_i,id_x}\right).
			\label{eq:op-protocol-details-final-aggregate-decoding}
		\end{IEEEeqnarray}
		When the consumption readings are represented at the selected fixed-point
		resolution, this value equals their exact aggregate. The protocol-attempt
		identifier is omitted from \(C_{t_i,D_w}\) because every successful attempt
		for the same reporting interval produces the same aggregate consumption
		value.
		
		\par\noindent\textbf{Failure Handling, Attempt Restart, and State Erasure:}	
		A protocol attempt is considered successful only if all expected AEAD
		noise-component records are successfully processed, the aggregator receives
		exactly one Paillier ciphertext from every meter in
		\(\mathcal{M}_{D_w}\), and the final aggregate ciphertext is produced for the
		same protocol context \((D_w,e,t_i,a)\). If any required record or ciphertext
		is missing or invalid when the attempt timeout expires, the aggregator aborts
		the entire attempt and does not compute a partial aggregate.
		
		If the interval execution is restarted, the aggregator assigns a new attempt
		identifier:
		\begin{IEEEeqnarray}{rCl}
			a'
			&\leftarrow&
			a+1.
			\label{eq:op-protocol-details-attempt-restart}
		\end{IEEEeqnarray}
		The restarted attempt uses the same participant, designated, and
		non-designated meter sets for interval \(t_i\), but all attempt-specific
		values are generated again. In particular, the meters derive new
		interval-specific keys using \(a'\), generate new nonces and fresh noise
		components, and produce new AEAD and Paillier ciphertexts. No key, nonce,
		noise component, or ciphertext from the aborted attempt is reused.
		
		After successful completion or abortion of an attempt, the smart meters
		securely erase all corresponding interval-specific keys, nonces, noise
		components, decrypted AEAD plaintexts, and cached ciphertexts. The aggregator
		also discards the individual contributions and routing state associated with
		that attempt after they are no longer required. The epoch-specific pairwise
		master keys remain stored until epoch \(e\) expires.
		
		\subsection{Correctness of Noise Cancellation and Aggregate Recovery}
		\label{sec:protocol-correctness}
		
		For a successful protocol attempt, the sum of the noisy and
		cancellation-adjusted readings of all participating smart meters is:
		\begin{IEEEeqnarray}{rCl}
			\sum_{id_x\in\mathcal{M}_{D_w}}
			nc_{t_i,a,id_x}
			&=&
			\sum_{id_k\in\mathcal{N}_{t_i}}
			\left(
			c_{t_i,id_k}
			+
			\sum_{j=1}^{K}
			s_{t_i,a,id_k}^{id_{sel_j}}
			\right)
			\notag\\
			&&+
			\sum_{j=1}^{K}
			\left(
			c_{t_i,id_{sel_j}}
			-
			\sum_{id_k\in\mathcal{N}_{t_i}}
			s_{t_i,a,id_k}^{id_{sel_j}}
			\right)
			\notag\\
			&=&
			\sum_{id_x\in\mathcal{M}_{D_w}}
			c_{t_i,id_x}
			+
			\sum_{id_k\in\mathcal{N}_{t_i}}
			\sum_{j=1}^{K}
			s_{t_i,a,id_k}^{id_{sel_j}}
			\notag\\
			&&-
			\sum_{j=1}^{K}
			\sum_{id_k\in\mathcal{N}_{t_i}}
			s_{t_i,a,id_k}^{id_{sel_j}}
			\notag\\
			&=&
			\sum_{id_x\in\mathcal{M}_{D_w}}
			c_{t_i,id_x}
			=
			C_{t_i,D_w}.
			\label{eq:op-protocol-correctness-noise-cancellation}
		\end{IEEEeqnarray}
		The two double sums contain the same noise components and therefore cancel
		exactly. Consequently, the aggregate of the noisy and cancellation-adjusted
		readings equals the aggregate of the actual consumption readings. By the
		Paillier homomorphic property and the no-wrap condition defined in
		\eqref{eq:op-protocol-details-aggregate-no-wraparound}, decrypting the final
		aggregate ciphertext recovers \(C_{t_i,D_w}\).
			
		\section{Evaluation}
		We evaluated our scheme from two key aspects: (1) performance and (2) privacy.
		For the performance evaluation, the computational, memory, and communication
		overheads of the proposed protocol are examined. For the privacy evaluation,
		normalized conditional entropy (NCE) and normalized root-mean-square error
		(NRMSE) are used to evaluate the protection of individual consumption
		readings under the adversarial model defined in Section
		\ref{sec:threat-model-assumptions}. A separate security analysis examines the
		adversarial view and collusion, the protection of protocol values, and privacy
		under partial meter exposure.
		
		\subsection{Performance Evaluation}
		We first conduct an analytical and experimental evaluation of the protocol's
		computational overhead. We further examine its memory requirements and
		analytically evaluate the communication overhead based on the protocol
		messages exchanged between the participating entities.
		
		In our experimental setup, the protocol is deployed on two distinct systems.
		The first system, simulating the smart meter, employs the Quick Emulator
		(QEMU) to replicate an Orange Pi One PC, which is equipped with a 1.2 GHz
		32-bit ARM Cortex-A7 processor and 1 GB of RAM. This emulated environment
		operates on Armbian, an Ubuntu-based operating system tailored for Internet
		of Things (IoT) applications. The second system, equipped with a 1.8 GHz
		Intel Core i7 processor (8 cores) and 16 GB of RAM, represents the aggregator
		and the utility provider. To assess protocol feasibility under memory
		constraints, the RAM available on the second machine is intentionally
		restricted to 1 GB. The second system operates on Debian 13.
		
		We implemented our scheme in Python using optimized cryptographic libraries.
		The implementation employs authenticated DHKEM based on X25519 and
		HKDF-SHA-256 for pairwise key establishment, HKDF-SHA-256 for key
		derivation, and ChaCha20-Poly1305 for authenticated encryption. We use
		LightPHE \cite{serengil2024lightphe} for the Paillier cryptosystem.
		Gaussian noise components are generated using an operating-system-backed
		cryptographically secure pseudorandom number generator (CSPRNG) together
		with the Box--Muller transformation.
		
		We utilized a dataset \cite{cenky2023dataset} from Slovakia's AMI infrastructure consisting of energy consumption data from 1000 anonymized residential smart meters. The dataset includes both active and reactive energy readings, recorded at 15-minute intervals over the course of a full year. This dataset is privatized using traditional techniques such as data minimization and anonymization. It is also important to note that it does not suffer from null values and data sparsity issues.
		
		\subsubsection{Computational Overhead}
		To analytically evaluate the protocol's computational overhead, we consider
		the main recurring cryptographic operations performed by each entity. As
		different parts of the protocol are executed by different entities, we
		examine the computational overhead for each entity separately. Table
		\ref{table:op-evaluation-computational-overhead-notation} summarizes the
		symbols used in this analysis.
		
		\begin{table}[htp]
			\centering
			\caption{Notation for Computational Overhead}
			\small
			\renewcommand{\arraystretch}{1.3}
			\setlength{\tabcolsep}{5pt}
			\resizebox{1.00\columnwidth}{!}{
				\begin{tabular}{|c|p{10cm}|}
					\hline
					\textbf{Symbol} & \textbf{Description} \\
					\hline
					$\mathbf{t}_{G}$ &
					Time of generating one Gaussian noise component \\
					\hline
					$\mathbf{t}_{KDF}$ &
					Time of one interval-key derivation \\
					\hline
					$\mathbf{t}_{AEAD}$ &
					Time of one AEAD encryption or decryption \\
					\hline
					$\mathbf{t}_{PEnc}$ &
					Time of one Paillier encryption \\
					\hline
					$\mathbf{t}_{PH}$ &
					Time of one Paillier homomorphic ciphertext combination \\
					\hline
					$\mathbf{t}_{PDec}$ &
					Time of one Paillier decryption \\
					\hline
				\end{tabular}
			}
			\label{table:op-evaluation-computational-overhead-notation}
		\end{table}
		
		A non-designated smart meter generates \(K\) Gaussian noise components,
		derives \(K\) interval-specific keys, performs \(K\) AEAD encryptions, and
		performs one Paillier encryption. Therefore, its computational overhead is:
		\begin{IEEEeqnarray}{rCl}
			\mathbf{t}_{sm}^{ND}
			&=&
			K\left(
			\mathbf{t}_{G}
			+\mathbf{t}_{KDF}
			+\mathbf{t}_{AEAD}
			\right)
			+\mathbf{t}_{PEnc}.
			\label{eq:op-evaluation-comp-nd}
		\end{IEEEeqnarray}
		
		Each designated smart meter processes one protected noise component from
		every non-designated meter. It therefore performs \(M-K\) interval-key
		derivations, \(M-K\) AEAD decryptions, and one Paillier encryption. Its
		computational overhead is:
		\begin{IEEEeqnarray}{rCl}
			\mathbf{t}_{sm}^{D}
			&=&
			(M-K)
			\left(
			\mathbf{t}_{KDF}
			+\mathbf{t}_{AEAD}
			\right)
			+\mathbf{t}_{PEnc}.
			\label{eq:op-evaluation-comp-d}
		\end{IEEEeqnarray}
		
		The aggregator homomorphically combines the \(M\) Paillier ciphertexts
		received from the participating smart meters. Hence, its main computational
		overhead is:
		\begin{IEEEeqnarray}{rCl}
			\mathbf{t}_{agg}
			&=&
			(M-1)\mathbf{t}_{PH}.
			\label{eq:op-evaluation-comp-agg}
		\end{IEEEeqnarray}
		
		The utility provider performs one Paillier decryption to recover the final
		aggregated consumption value. Therefore:
		\begin{IEEEeqnarray}{rCl}
			\mathbf{t}_{up}
			&=&
			\mathbf{t}_{PDec}.
			\label{eq:op-evaluation-comp-up}
		\end{IEEEeqnarray}
		
		Accordingly, the aggregate computational workload of one successful
		reporting attempt is:
		\begin{IEEEeqnarray}{rCl}
			\mathbf{t}_{op}
			&=&
			(M-K)\mathbf{t}_{sm}^{ND}
			+K\mathbf{t}_{sm}^{D}
			+\mathbf{t}_{agg}
			+\mathbf{t}_{up}.
			\label{eq:op-evaluation-comp-total}
		\end{IEEEeqnarray}
		
		We also experimentally evaluated the protocol's computational performance.
		The experiments use 20 participating smart meters and
		\(K\in\{3,5,10,15\}\), while the Paillier modulus size is varied between
		1024, 2048, and 3072 bits. For each configuration, the protocol is executed
		100 times after three warm-up executions, and Table
		\ref{table:op-service-evaluation} reports the mean execution time of each
		entity. The 1024-bit Paillier configuration is included only as a performance
		baseline.
		
		\begin{table}[htp]
			\centering
			\caption{Mean Protocol Execution Time per Entity (seconds)}
			\scriptsize
			\renewcommand{\arraystretch}{1.15}
			\setlength{\tabcolsep}{3pt}
			\resizebox{1.00\columnwidth}{!}{
				\begin{tabular}{|c|c|c|c|c|c|c|}
					\hline
					\textbf{Paillier} &
					\(\mathbf{K}\) &
					\textbf{Designated SM} &
					\textbf{Non-Designated SM} &
					\textbf{AGG} &
					\textbf{UP} &
					\textbf{Total Execution Time} \\
					\hline
					
					1024 & 3  & 0.183367 & 0.240505 & 0.000222 & 0.011378 & 0.435472 \\
					1024 & 5  & 0.262914 & 0.209317 & 0.000233 & 0.011982 & 0.484446 \\
					1024 & 10 & 0.214781 & 0.211677 & 0.000221 & 0.011397 & 0.438076 \\
					1024 & 15 & 0.220955 & 0.272012 & 0.000227 & 0.011705 & 0.504899 \\
					\hline
					
					2048 & 3  & 1.064615 & 1.099932 & 0.000760 & 0.077322 & 2.242629 \\
					2048 & 5  & 1.273349 & 1.169397 & 0.000767 & 0.078316 & 2.521829 \\
					2048 & 10 & 1.166824 & 1.087126 & 0.000749 & 0.076289 & 2.330988 \\
					2048 & 15 & 0.973643 & 1.268198 & 0.000753 & 0.076664 & 2.319258 \\
					\hline
					
					3072 & 3  & 3.472572 & 3.603712 & 0.001535 & 0.242822 & 7.320641 \\
					3072 & 5  & 3.772078 & 3.455227 & 0.001546 & 0.244657 & 7.473508 \\
					3072 & 10 & 3.059876 & 3.512165 & 0.001545 & 0.244389 & 6.817975 \\
					3072 & 15 & 3.624267 & 2.847408 & 0.001544 & 0.244970 & 6.718189 \\
					\hline
				\end{tabular}
			}
			\label{table:op-service-evaluation}
		\end{table}

		As shown in Table \ref{table:op-service-evaluation}, the Paillier modulus
		size has the dominant effect on the execution time. Increasing \(K\) mainly
		changes the number of lightweight Gaussian, KDF, and AEAD operations, whereas
		each participating smart meter still performs only one Paillier encryption.
		The aggregator maintains a very small computational overhead because it only
		combines the received Paillier ciphertexts, while the utility provider
		performs a single Paillier decryption.
		
		\subsubsection{Memory Overhead}
		We experimentally evaluated the protocol's memory requirement for all tested
		values of \(K\) and Paillier modulus sizes. Across all configurations, The maximum process memory consumption was approximately 58.81 MB on the meter-simulator system and 90.22 MB on the aggregator-simulator system. These values are significantly lower than the available 1 GB of RAM, demonstrating that the protocol can operate within the considered memory constraint.
		
		\subsubsection{Communication Overhead}
		Lastly, we assess the protocol's communication overhead.
		
		In this study, we assess the communication overhead in both Neighborhood
		Area Networks (NAN) and Wide Area Networks (WAN) by considering the link
		layer technologies employed. In our network architecture, smart meters
		transmit data to an aggregator using 6LoWPAN alongside IEEE 802.15.4g
		(Wi-SUN)---a standard specifically designed for NAN that connects smart
		meters to intermediary local aggregators \cite{chang2012ieee}. The local
		aggregator aggregates the received encrypted consumption contributions and
		then relays the encrypted aggregate to the utility provider via a 4G-LTE
		network. More specifically, the data is initially sent to the 4G-LTE Radio
		Access Network (RAN), also known as the Evolved Node B (eNB), then forwarded
		to the Packet Data Network Gateway (PGW), and finally delivered to the
		utility provider.
		
		Table \ref{table:op-communication-overhead} reports the communication
		overhead handled by one entity in each recurring protocol exchange. For the
		NAN, 12.5 kbps is allocated to each smart meter from the available
		250 kbps Wi-SUN bandwidth. For the AGG-to-eNB link, 50 kbps is allocated
		to each aggregator because multiple aggregators may serve different
		districts and share the available 1000 kbps WAN bandwidth. After the
		traffic reaches the eNB, the eNB-to-PGW and PGW-to-UP links are evaluated
		using the full 1000 kbps link bandwidth.
		
		The ``Messages per entity'' column specifies how many separate application
		messages are handled by the entity represented in each row. In particular,
		during noise-record routing, one designated smart meter receives one
		separate record from each of the \(M-K\) non-designated smart meters.
		Therefore, the application bytes, transmitted bytes, number of frames, and
		transmission time reported for this exchange are the cumulative values
		handled by one designated smart meter.
		
		\begin{table*}[htp]
			\centering
			\caption{Per-Entity Communication Overhead
				(\(M=20\), 3072-bit Paillier)}
			\scriptsize
			\renewcommand{\arraystretch}{1.25}
			\setlength{\tabcolsep}{3pt}
			
			\resizebox{\textwidth}{!}{%
				\begin{tabular}{|
						>{\centering\arraybackslash}p{6.2cm}|
						c|c|c|
						c|c|
						c|c|
						c|c|
						c|c|}
					\hline
					
					\multirow{3}{*}{\textbf{Protocol Message}} &
					\multirow{3}{*}{\(\mathbf{K}\)} &
					\multirow{3}{*}{\textbf{Messages/entity}} &
					\multirow{3}{*}{\textbf{Application bytes}} &
					\multicolumn{2}{c|}{\textbf{NAN}} &
					\multicolumn{6}{c|}{\textbf{WAN}} \\ \cline{5-12}
					
					& & & &
					\multicolumn{2}{c|}{%
						\begin{tabular}[c]{@{}c@{}}
							\textbf{SM $\leftrightarrow$ AGG}\\
							IEEE 802.15.4g (Wi-SUN)
					\end{tabular}}
					& \multicolumn{2}{c|}{%
						\begin{tabular}[c]{@{}c@{}}
							\textbf{AGG $\rightarrow$ eNB}\\
							4G-LTE (PDCP-LTE)
					\end{tabular}}
					& \multicolumn{2}{c|}{%
						\begin{tabular}[c]{@{}c@{}}
							\textbf{eNB $\rightarrow$ PGW}\\
							IEEE 802.3 (Ethernet)
					\end{tabular}}
					& \multicolumn{2}{c|}{%
						\begin{tabular}[c]{@{}c@{}}
							\textbf{PGW $\rightarrow$ UP}\\
							IEEE 802.3 (Ethernet)
					\end{tabular}} \\ \cline{5-12}
					
					& & & &
					\textbf{Tx Size (Frames)} & \textbf{Time (s)} &
					\textbf{Tx Size} & \textbf{Time (s)} &
					\textbf{Tx Size} & \textbf{Time (s)} &
					\textbf{Tx Size} & \textbf{Time (s)} \\
					\hline
					
					\multirow{4}{6.2cm}{\centering
						\(\begin{aligned}
							AGG \rightarrow SMs:\;&
							D_w\Vert e\Vert t_i\Vert a
							\Vert\mathcal{D}_{t_i}
						\end{aligned}\)}
					& 3 & 1 & 64 B
					& 89 B (1) & 0.05696
					& -- & -- & -- & -- & -- & -- \\
					
					& 5 & 1 & 84 B
					& 109 B (1) & 0.06976
					& -- & -- & -- & -- & -- & -- \\
					
					& 10 & 1 & 134 B
					& 193 B (2) & 0.12352
					& -- & -- & -- & -- & -- & -- \\
					
					& 15 & 1 & 184 B
					& 243 B (2) & 0.15552
					& -- & -- & -- & -- & -- & -- \\
					\hline
					
					\multirow{4}{6.2cm}{\centering
						\(\begin{aligned}
							SM_k \rightarrow AGG:\;&
							D_w\Vert e\Vert t_i\Vert a\Vert id_k
							\Vert\mathcal{X}_{UP,t_i,a}^{k}\\
							&\Vert\mathcal{R}_{t_i,a,k,1}
							\Vert\cdots\Vert
							\mathcal{R}_{t_i,a,k,K}
						\end{aligned}\)}
					& 3 & 1 & 974 B
					& 1,273 B (10) & 0.81472
					& -- & -- & -- & -- & -- & -- \\
					
					& 5 & 1 & 1,082 B
					& 1,411 B (11) & 0.90304
					& -- & -- & -- & -- & -- & -- \\
					
					& 10 & 1 & 1,352 B
					& 1,771 B (14) & 1.13344
					& -- & -- & -- & -- & -- & -- \\
					
					& 15 & 1 & 1,622 B
					& 2,101 B (16) & 1.34464
					& -- & -- & -- & -- & -- & -- \\
					\hline
					
					\multirow{4}{6.2cm}{\centering
						\(\begin{aligned}
							AGG \rightarrow SM_{sel_j}:\;&
							D_w\Vert e\Vert t_i\Vert a\Vert id_k
							\Vert\mathcal{R}_{t_i,a,k,j}
						\end{aligned}\)}
					& 3 & 17 & 1,666 B
					& 2,091 B (17) & 1.33824
					& -- & -- & -- & -- & -- & -- \\
					
					& 5 & 15 & 1,470 B
					& 1,845 B (15) & 1.18080
					& -- & -- & -- & -- & -- & -- \\
					
					& 10 & 10 & 980 B
					& 1,230 B (10) & 0.78720
					& -- & -- & -- & -- & -- & -- \\
					
					& 15 & 5 & 490 B
					& 615 B (5) & 0.39360
					& -- & -- & -- & -- & -- & -- \\
					\hline
					
					\(\begin{aligned}
						SM_{sel_j} \rightarrow AGG:\;&
						D_w\Vert e\Vert t_i\Vert a\Vert id_{sel_j}
						\Vert\mathcal{X}_{UP,t_i,a}^{sel_j}
					\end{aligned}\)
					& All & 1 & 812 B
					& 1,051 B (8) & 0.67264
					& -- & -- & -- & -- & -- & -- \\
					\hline
					
					\(\begin{aligned}
						AGG \rightarrow UP:\;&
						D_w\Vert e\Vert t_i\Vert a
						\Vert\mathscr{X}_{t_i,a,D_w}
					\end{aligned}\)
					& All & 1 & 802 B
					& -- & --
					& 856 B & 0.13696
					& 924 B & 0.007392
					& 868 B & 0.006944 \\
					\hline
					
					\multicolumn{4}{|c|}{\textbf{Available BW}}
					& \multicolumn{2}{c|}{250 kbps}
					& \multicolumn{2}{c|}{1000 kbps}
					& \multicolumn{2}{c|}{\multirow{2}{*}{1000 kbps}}
					& \multicolumn{2}{c|}{\multirow{2}{*}{1000 kbps}} \\
					\cline{1-8}
					
					\multicolumn{4}{|c|}{\textbf{Allocated BW}}
					& \multicolumn{2}{c|}{12.5 kbps (per meter)}
					& \multicolumn{2}{c|}{50 kbps (per AGG)}
					& \multicolumn{2}{c|}{}
					& \multicolumn{2}{c|}{} \\
					\hline
					
				\end{tabular}%
			}
			\label{table:op-communication-overhead}
		\end{table*}
		
		Table \ref{table:op-communication-overhead} shows that the communication
		requirements of the two smart-meter roles change differently with \(K\).
		For a non-designated smart meter, increasing \(K\) increases the size of its
		submission because the meter sends one AEAD-protected noise record for each
		designated smart meter.
		
		For a designated smart meter, the opposite behavior is observed during
		noise-record routing. Each designated smart meter receives one separate
		record from every non-designated smart meter, giving \(M-K\) received
		messages. Consequently, the number of routed messages handled by one
		designated smart meter decreases. Finally, after receiving all meter contributions, the aggregator sends only
		one encrypted aggregate toward the utility provider. 
		
		\subsection{Privacy Evaluation}
		To evaluate the privacy provided by the proposed scheme, we use two
		complementary metrics: normalized conditional entropy (NCE) and normalized
		root-mean-square error (NRMSE). NCE measures the uncertainty that remains
		about the actual consumption values after observing their perturbed protocol
		values, whereas NRMSE measures the accuracy of adversarial reconstruction
		under partial meter exposure. The same consumption data are used for all
		configurations, and each result is averaged over 100 independent noise
		realizations. For NCE calculation, the same discretization ranges derived from the actual
		consumption readings are used for all configurations.
		
		Let \(X\) denote the actual consumption values and \(Y\) the corresponding
		perturbed observation. Following \cite{wagner2018technical}, NCE is defined
		as:
		\begin{IEEEeqnarray}{rCl}
			H_{CE}(X|Y)
			&=&
			-\sum_{x,y}p(x,y)\log_2 p(x|y),
			\notag\\
			H_E(X)
			&=&
			-\sum_x p(x)\log_2 p(x),
			\notag\\
			H_{NCE}(X|Y)
			&=&
			\frac{H_{CE}(X|Y)}{H_E(X)}.
			\label{eq:op-protocol-evaluation-privacy-nce}
		\end{IEEEeqnarray}
		NCE lies in \([0,1]\), where a value closer to 1 indicates that the
		perturbed observation leaves greater uncertainty about the actual
		consumption values.
		
		We evaluate both smart-meter roles using the noise-scale factor
		\(\alpha\in\{1/9,1/6,1/3,1,3,6,9\}\). Since each noise component has
		standard deviation \(\alpha\sigma/\sqrt{K}\), the two observable values
		satisfy:
		\begin{IEEEeqnarray}{rCl}
			Y_{ND}
			&=&
			X+\sum_{j=1}^{K}S_j,
			\qquad
			\operatorname{Var}(Y_{ND}-X)=\alpha^2\sigma^2,
			\notag\\
			Y_D
			&=&
			X-\sum_{k=1}^{M-K}S_k,
			\qquad
			\operatorname{Var}(Y_D-X)
			=
			\frac{M-K}{K}\alpha^2\sigma^2. \notag \\ 
			\label{eq:op-protocol-evaluation-privacy-observations}
		\end{IEEEeqnarray}
		Thus, the total perturbation of a non-designated target is independent of
		\(K\), whereas the perturbation of a designated target depends on
		\((M-K)/K\).
		
		\begin{figure*}[htp]
			\centering
			\includegraphics[width=0.92\textwidth]
			{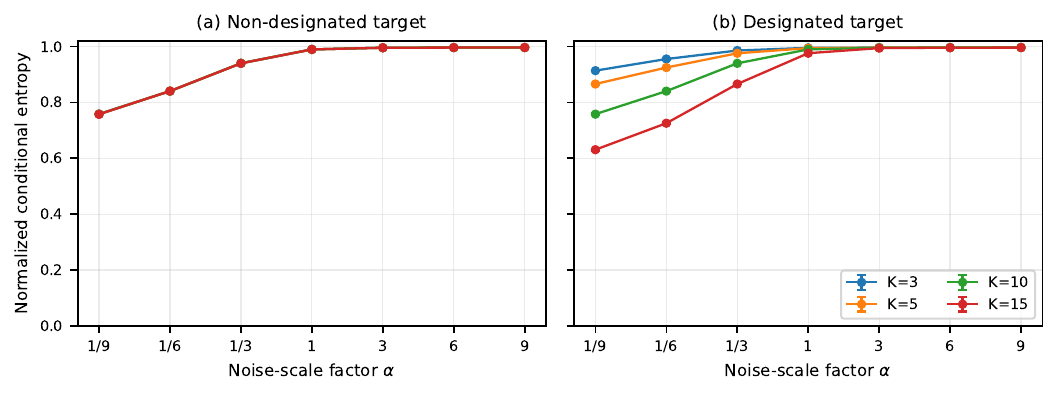}
			\caption{NCE under different noise-scale factors for (a) an unexposed
				non-designated target and (b) an unexposed designated target.
				Higher NCE indicates greater uncertainty about the actual consumption
				values. Error bars show the standard deviation over 100 trials.}
			\label{op-protocol-privacy-eval-NCE}
		\end{figure*}
		
		Fig. \ref{op-protocol-privacy-eval-NCE} shows that NCE increases with the
		noise scale for both roles. The non-designated curves are nearly identical
		for all values of \(K\), as expected from their constant total noise
		variance. At \(\alpha=1\), their mean NCE is approximately \(0.99\).
		For designated targets, \(K\) has a stronger effect at small noise scales
		because their perturbation variance depends on \((M-K)/K\). As the noise
		scale increases, all configurations approach an NCE value close to 1.
		
		To evaluate privacy when some meters are exposed, we use range-normalized
		RMSE:
		\begin{IEEEeqnarray}{rCl}
			NRMSE
			&=&
			\frac{
				\sqrt{\frac{1}{N}
					\sum_{i=1}^{N}(\widehat{c}_i-c_i)^2}}
			{\max(X)-\min(X)}.
			\label{eq:op-protocol-evaluation-privacy-nrmse}
		\end{IEEEeqnarray}
		A higher NRMSE represents a less accurate adversarial reconstruction.
		Unlike NCE, NRMSE is not restricted to \([0,1]\).
		
		For an unexposed non-designated target, exposure of \(r_D\) designated
		meters reveals \(r_D\) of its noise components. For an unexposed designated
		target, exposure of \(r_N\) non-designated meters reveals the corresponding
		components of its cancellation value. The resulting reconstruction errors
		are characterized by:
		\begin{IEEEeqnarray}{rCl}
			\widehat{X}_{ND}^{(r_D)}
			&=&
			Y_{ND}-\sum_{j\in\mathcal{E}_D}S_j,
			\qquad
			\sigma_{res,ND}
			=
			\sigma\sqrt{\frac{K-r_D}{K}},
			\notag\\
			\widehat{X}_{D}^{(r_N)}
			&=&
			Y_D+\sum_{k\in\mathcal{E}_N}S_k,
			\qquad
			\sigma_{res,D}
			=
			\sigma\sqrt{\frac{M-K-r_N}{K}}. \notag \\ 
			\label{eq:op-protocol-evaluation-privacy-reconstruction}
		\end{IEEEeqnarray}
		
		\begin{figure*}[htp]
			\centering
			\includegraphics[width=0.92\textwidth]
			{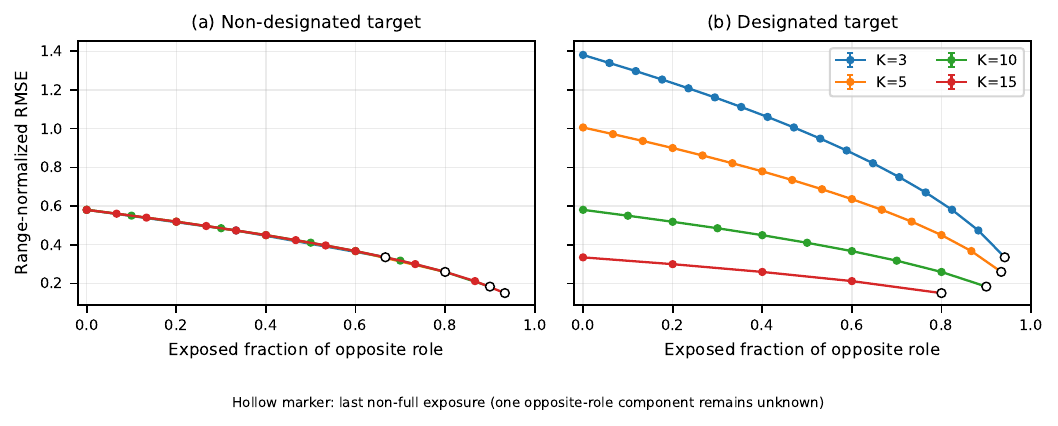}
			\caption{NRMSE under partial meter exposure for (a) an unexposed
				non-designated target and (b) an unexposed designated target.
				The exposed fraction is relative to the opposite role. Hollow markers
				indicate the last non-full exposure, where one noise component remains
				unknown. Error bars show the standard deviation over 100 trials.}
			\label{op-protocol-privacy-eval-NRMSE}
		\end{figure*}
		
		Fig. \ref{op-protocol-privacy-eval-NRMSE} shows that NRMSE decreases as
		more opposite-role meters are exposed because the adversary learns and
		removes additional noise components. For non-designated targets, the
		zero-exposure NRMSE is approximately \(0.58\) for all tested \(K\), which
		is consistent with their \(K\)-independent total noise variance. For
		designated targets, the initial NRMSE depends more strongly on \(K\) because
		their residual noise depends on \((M-K)/K\). The hollow markers show the
		last non-full exposure case; one noise component remains unknown and exact
		reconstruction is therefore still prevented under the evaluated model,
		although the reconstruction error has substantially decreased.
		
		\subsection{Security Analysis}
		
		\textbf{Adversarial View and Protection:}
		We consider the adversarial view defined in Section
		\ref{sec:threat-model-assumptions}, where the aggregator and utility provider
		may collude and combine their protocol views with the internal states of a
		set \(\mathcal{E}\) of exposed smart meters. The aggregator can see the designated-meter set, the Paillier ciphertexts	submitted by the smart meters, and the AEAD-protected noise records that it forwards. If the aggregator colludes with the utility provider, the utility provider can use its Paillier secret key to decrypt the submitted meter	contributions. However, these plaintext contributions are still noisy or cancellation-adjusted values rather than the exact individual readings.
		
		The noise components themselves remain protected between the corresponding
		non-designated and designated meters. They are encrypted using AEAD keys
		derived from pairwise master keys that are not known to the aggregator or the
		utility provider. If a meter is exposed, the coalition learns only the keys
		and noise values associated with that exposed meter. The pairwise secrets
		shared only between unexposed meters remain unknown. Therefore, the coalition
		can cancel only the noise components revealed through exposed meters, while
		the remaining unknown components continue to hide the exact reading.
		
		\textbf{Privacy under Partial Meter Exposure:}
		Consider an unexposed non-designated meter \(id_k\). Even if the coalition
		decrypts its Paillier contribution, it obtains only
		\begin{IEEEeqnarray}{rCl}
			nc_{t_i,a,id_k}
			&=&
			c_{t_i,id_k}
			+
			\sum_{id_{sel_j}\in\mathcal{D}_{t_i}}
			s_{t_i,a,id_k}^{id_{sel_j}}.
			\label{eq:security-nd-view}
		\end{IEEEeqnarray}
		Noise components associated with exposed designated meters may be known and
		removed. However, if
		\(\mathcal{D}_{t_i}\setminus\mathcal{E}\neq\varnothing\), at least one
		independent component remains unknown, preventing exact recovery of
		\(c_{t_i,id_k}\).
		
		Similarly, for an unexposed designated meter \(id_{sel_j}\), the coalition
		may obtain its cancellation-adjusted contribution
		\begin{IEEEeqnarray}{rCl}
			nc_{t_i,a,id_{sel_j}}
			&=&
			c_{t_i,id_{sel_j}}
			-
			\sum_{id_k\in\mathcal{N}_{t_i}}
			s_{t_i,a,id_k}^{id_{sel_j}}.
			\label{eq:security-d-view}
		\end{IEEEeqnarray}
		If
		\(\mathcal{N}_{t_i}\setminus\mathcal{E}\neq\varnothing\), at least one
		noise component generated by an unexposed non-designated meter remains
		unknown. Hence, the exact designated-meter reading cannot be recovered.
		
		Since
		\(|\mathcal{D}_{t_i}|=K\),
		\(|\mathcal{N}_{t_i}|=M-K\), and
		\(|\mathcal{E}|\leq f\), the sufficient exposure condition
		\begin{IEEEeqnarray}{rCl}
			f
			&<&
			\min(K,M-K)
			\notag\\
			&\Longrightarrow&
			\mathcal{D}_{t_i}\setminus\mathcal{E}\neq\varnothing
			\;\land\;
			\mathcal{N}_{t_i}\setminus\mathcal{E}\neq\varnothing
			\label{eq:security-exposure-bound}
		\end{IEEEeqnarray}
		ensures that at least one unexposed meter remains in each role. Therefore,
		under the stated adversarial model, the coalition cannot exactly recover
		the individual reading of a trusted and unexposed smart meter.
		
		\section{Conclusion}
		In this paper, we proposed a collusion-resistant privacy-preserving smart
		metering protocol for operational utility services. The protocol distributes
		noise-cancellation responsibility among \(K\) designated smart meters and
		combines Paillier homomorphic aggregation with a KEM--KDF--AEAD construction
		for protecting individual noise components. The generated noise cancels
		during aggregation, allowing the utility provider to recover the exact
		aggregate while individual meter contributions remain perturbed.
		
		Our evaluation shows that the protocol is feasible for the considered
		smart-meter environment and that its privacy behavior follows the intended
		design. NCE increases with the applied noise scale, while NRMSE shows that
		adversarial reconstruction becomes more accurate as additional
		opposite-role meters are exposed. Finally, the security analysis shows that
		under
		\(f<\min(K,M-K)\), at least one designated and one non-designated meter
		remain unexposed, leaving an independent noise component unknown to the
		adversarial coalition and preventing exact recovery of the reading of a
		trusted and unexposed smart meter.

		
		
		%
		
		



		\ifCLASSOPTIONcaptionsoff
		\newpage
		\fi

		
		
		
		
		\bibliographystyle{IEEEtran}
		\bibliography{IEEEabrv,Bibliography}

		\vfill

	\end{document}